\documentstyle [aps]{revtex}
\begin{document}
\title{NONLINEAR QUANTUM EVOLUTION WITH\\ MAXIMAL ENTROPY PRODUCTION}
\author{S. Gheorghiu-Svirschevski\footnotemark[1]\footnotetext{e-mail: hnmg@soa.com}}
\address{1087 Beacon St., Suite 301, Newton, MA 02459}
\date{\today}
\maketitle
\begin{abstract}
We derive a well-behaved nonlinear extension of the
non-relativistic Liouville-von Neumann dynamics driven by maximal
entropy production with conservation of energy and probability.
The pure state limit reduces to the usual Schroedinger evolution,
while mixtures evolve towards maximum entropy equilibrium states with
canonical-like probability distributions on energy eigenstates.
The linear, near-equilibrium limit is found to amount to an
essentially exponential relaxation to thermal equilibrium; a few
elementary examples are given. In addition, the modified dynamics
is invariant under the time-independent symmetry group of the
hamiltonian, and also invariant under the special Galilei group
provided the conservation of total momentum is accounted for as
well. Similar extensions can be generated for, e.g., nonextensive systems
better described by a Tsallis q-entropy.
\end{abstract}
\pacs{11.10Lm, 03.65Bz, 04.60-m, 05.45-a}

%\narrowtext
%\twocolumn
\section{INTRODUCTION}
\label{sec1} A number of recent, independent experiments \cite{1}
have provided impressive bounds on the possible deviations from a
linear and unitary propagation of pure quantum states, at least
on a laboratory accessible space-time scale. The limits imposed
in this way on potential generalizations of the standard unitary
quantum equations of motions, as sought in relation to Hawking's
blackhole evaporation process \cite{2}, are likewise severe.
Certainly, there always remains the possibility of modified
dynamical laws on the (inaccessible) Plank scale \cite{3}, as well
as under the extreme physical environment characteristic of
singular cosmological phenomena. Related models of open system
dynamics due to alleged statistical perturbations, e.g. from the
space-time foam, have enjoyed considerable attention lately
\cite{4}. But in case the unitarity of pure state propagation
holds under universal conditions, one is necessarily lead to a
quest for genuine nonlinear extensions for isolated systems,
possibly involving an explicit arrow of time. Indeed, it has been
pointed out in a fairly general ansatz \cite{5,6} that if the pure
states happen to be attractors of a nonlinear evolution, then
testing the unitary propagation of pure states alone cannot rule
out a nonlinear propagation of mixtures. This situation has been noted recently in the context of certain nonlinear Lie-Poisson dynamics \cite{6}, wherein pure states still propagate in the usual hamiltonian way, while density matrices evolve nonlinearly, but preserving a time-independent spectrum. Unfortunately, the underlying physics remains rather obscure in these theories and the selection of particular realizations relevant to various experimental setups is, in general, a matter of guesswork.

In the following we show that a physically meaningful nonlinear
extension emerges when the fundamental postulates of quantum
mechanics are supplemented by the first and second principles of
thermodynamics, at the sole expense of ignoring the constraint of
a linear, unitary evolution in time. The result is a largely
irreversible, highly nonlinear generalization of the
non-relativistic quantum Liouville equation, of a form closely
related to the ansatz of Ref.\onlinecite{5}(but not in the
Lie-Poisson class), which features a number of rather intriguing
properties. In particular, pure states are still propagated
unitarily into pure states according to the usual
(time-reversible) hamiltonian dynamics. The same is true of mixed
states characterized by an initial equiprobable distribution on a
(finite) set of uncorrelated (orthogonal) states. Non-pure states evolve so as to maximize the entropy production at each moment in time and to reach stationary states of maximum entropy (or minimum entropy production, according to Prigogine's nonequilibrium principle \cite{7}) on the shortest path in the appropriate state space. Precisely, mixed states arbitrarily distributed on a finite set of uncorrelated states
evolve into mixed states distributed on an equal number of
uncorrelated states, have a time-dependent eigenspectrum and eventually attain stationarity on a
subset of energy eigenstates. A similar statement can be
inferred, by extension, for mixtures of an infinite set of
uncorrelated pure states. It follows as well that the probability
distribution at equilibrium, on (a subset of) energy eigenstates,
has a canonical-like dependence on the energy eigenvalues. For
mixtures with an infinite energy range, the corresponding
temperature is, of course, strictly positive, whereas for
mixtures of a finite set of pure states the stationary state may
display a ``negative-temperature'' distribution, in analogy to
systems with a finite-dimensional state space. The above
mentioned properties are endorsed by the positivity of the
underlying evolution equation, which ensues by construction
despite the high degree of nonlinearity involved. The nature of
this essentially irreversible propagation becomes evident in the
close-to-equilibrium limit, when the matrix elements of the
density operator between energy eigenstates are found to undergo
simple exponential decays to the canonical equilibrium values.
Finally, proper (non-relativistic) invariance and conservation
properties under the symmetry group of the hamiltonian are also
accounted for. However, in the absence of an explicit general law
of entropy increase, the time scale for thermal relaxation is set
by one multiplication factor, a scalar functional, which is yet
to be given a specific expression.

Unlike the nonlinear Lie-Poisson dynamics \cite{6}, our framework
apparently challenges the notion of separability of isolated,
non-interacting systems, lack of which has long been thought to
be unacceptable \cite{8}. We argue, nevertheless, that in a
nonlinear theory it is necessary to refine the operational
definition of isolation and to acknowledge that the mutual
isolation of two non-interacting systems prohibits entanglement,
if individual time-translation invariance is to be preserved.
When this restriction is properly taken into account in the
formulation of the corresponding equation of motion, separability
can be easily recovered. On the other hand, the case where
non-interacting subsystems are allowed to develop correlations
spontaneously and eventually exchange energy (heat) is shown to
correspond in our ansatz to the phenomenon of ideal thermal
contact. From a precise technical perspective, the effect has its
origin in that the second principle applies, as usual, to the
total entropy of a compound system and not to the entropies of
individual subsystems. This necessarily results in such
redistribution of probabilities and energy as to maximize the
overall entropy. In physical terms, an ideal gas is allowed to
relax spontaneously to thermal equilibrium.

The formalism can be adapted straightforwardly to cover nonstandard forms for the entropy and energy functionals. As immediate examples, we construct a generalization of the Lie-Poisson dynamics with maximal entropy production and a nonlinear extension of the standard von Neumann evolution with maximal increase of the nonextensive Tsallis q-entropy \cite{9}.

\section{THE MODIFIED EQUATION OF MOTION}
\label{sec2} 

Following an earlier suggestion \cite{10}, the state
of a quantum system will be represented by a generalized
``square-root'' $\gamma$ of the density matrix $\rho$, defined by

\begin{equation}
\label{eq1}
\rho = \gamma \gamma ^{ +}\;.
\end{equation}

\noindent
In analogy to the common terminology, the operator
$\gamma$ (not necessarily hermitian) will be called here a state
operator. Note that the above decomposition is always well-defined, although not unique, for any hermitian and positive definite $\rho$. On the other hand, to any given $\gamma$ there corresponds a unique hermitian and positively defined $\rho$. We also adopt the standard inner product on the
associated Hilbert space of operators,

\begin{equation}
\label{eq2}
(\beta |\gamma) = Tr(\beta ^{ +} \gamma)\;,
\end{equation}

\noindent
such that for $\gamma $ normalized, $(\gamma |\gamma) =
Tr( \gamma ^{ +} \gamma) = 1$, the average of an observable
\textit{O} becomes the bilinear form

\begin{equation}
\label{eq3}
(\gamma |\textbf{O}|\gamma) = (\gamma ^{+} O\gamma) =
Tr(O\rho)\;,
\end{equation}

\noindent
with \textbf{O} the super-operator defined by
\textit{O},

\begin{equation}
\label{eq4}
\textbf{O}|\gamma) = |O\gamma)\;.
\end{equation}

It is further convenient to define the tilde-conjugate ${\tilde
\textbf {A}}$ of an arbitrary, and not necessarily linear,
super-operator \textbf{A} \cite{11}, by

\begin{equation}
\label{eq5}
\left(\textbf{A}|\alpha)\right)^{+}=\tilde{\textbf{A}}|\alpha^{+})\;.
\end{equation}

\noindent
It can be immediately verified that the super-operator
\textbf{A} maps hermitian operators $\alpha = \alpha ^{ +} $ into
hermitian operators $\beta = \beta ^{ +}  = \textbf{A}|\alpha)$ if
and only if it is tilde-symmetric, $\textbf{A}={\tilde \textbf
{A}}$. For a super-operator generated by a linear operator, such
as in Eq (\ref{eq4}) above, the tilde-conjugate is given by

\begin{equation}
\label{eq6}
\tilde{\textbf{A}}|\alpha)= |\alpha A^{ +})\;.
\end{equation}

\noindent
In particular, for the hermitian observable \textit{O}
it reads

\begin{equation}
\label{eq7}
\tilde{\textbf{O}}|\alpha) = |\alpha O)\;.
\end{equation}

\noindent
The tilde operation is distributive against the addition
and multiplication of super-operators, $\widetilde{(\textbf{A+B}
)} = \tilde{\textbf{A}} + \tilde{\textbf{B}}$,
$\widetilde{\textbf{AB}} = \tilde{\textbf{A}}\tilde{\textbf{B}}$,
and is anti-linear against multiplication by scalars,
$\widetilde{(a\textbf{A})} = a^*\tilde{\textbf{A}}$.

\bigskip
Let us consider now a massive isolated system characterized by an
energy operator (hamiltonian) \textit{H} and a state operator
$\gamma $ (density matrix $\rho=\gamma\gamma^{+}$), in an
inertial reference system where its center-of-mass is at rest. We
wish to find an equation of motion for this system which is
first-order differential in time and such that :

\noindent
1) \textit{Probability is conserved}:

\begin{equation}
\label{eq8}
\frac{d}{dt}(\gamma |\gamma) = \frac{d}{dt}Tr(\rho) =
0
\end{equation}

\noindent
or

\begin{equation}
(\dot{\gamma}|\gamma) + (\gamma|\dot{\gamma}) = 0 ,\quad
\dot{\gamma} = \frac{d}{dt}\gamma \;.\eqnum{8a}\label{eq8a}
\end{equation}

\noindent
2) \textit{Energy is conserved (first principle of
thermodynamics)}:

\begin{equation}
\label{eq9}
\frac{d}{dt}(\gamma|\textbf{H}|\gamma) =
\frac{d}{dt}Tr(H\rho)= 0
\end{equation}

\noindent
or

\begin{equation}
(\dot{\gamma}|\textbf{H}|\gamma) + (\gamma
|\textbf{H}|\dot{\gamma})= 0\;.\eqnum{9a}\label{eq9a}
\end{equation}

\noindent
3) \textit{The entropy production is always positive
(second principle of thermodynamics in non-equilibrium form)},

\begin{equation}
\label{eq10}
\frac{d}{dt}\textrm{S}(\textrm{t}) \geq 0\;.
\end{equation}

\noindent or

\begin{eqnarray}
\dot {\textrm{S}}(\textrm{t}) = -
[Tr(\dot{\rho}\ln\rho)+Tr(\dot{\rho})]
= - [(\;\dot{\gamma}\;|\;\ln(\gamma\gamma^{+})\;|\;\gamma)+%
(\;\gamma|\;\ln(\gamma\gamma^{+})\;|\;\dot{\gamma})+\nonumber\\
+(\;\dot{\gamma}|\;\gamma)+(\;\gamma|\;\dot{\gamma})] \geq 0\;.
\eqnum{10a}\label{eq10a}
\end{eqnarray}

\noindent where we adopt the standard entropy expression for a
normalized state ($Tr(\rho) = 1$)

\begin{equation}
\label{eq11}
\textrm{S}(\textrm{t}) = - \textrm{k}_{B}Tr\left[
\rho(\textrm{t})\;\ln\rho(\textrm{t})\right]
= - \textrm{k}_{B}(\;\gamma(\textrm{t})\;|\;\ln\left[\gamma(\textrm{t})\gamma^{+}%
(\textrm{t})\right]%
\;|\;\gamma(\textrm{t})\;)\;,
\end{equation}

\noindent with k$_{B}$ the Boltzmann constant.

In order to construct the desired equation of motion, we find it
convenient to consider a stronger form of the second principle,
by requiring that the entropy, as a functional of \textit{$\gamma
$}, increase in time along a path of maximum ascent. In other
words, let the entropy production (10a) be maximized, for any
given state $\gamma$, against variations of the time derivative
$\dot{\gamma}$, under the constraints (8a) of conservation of
probability and (9a) of conservation of energy. Note that the
variation of $\dot{\gamma}$ must avoid the simple multiplication
by a positive scalar, i.e. a trivial norm increase, since
$\dot{\textrm{S}}(\textrm{t})$ increases then unconditionally.
Hence the entropy production must be maximized against the
``direction'' of $\dot{\gamma}$, that is, against derivatives
$\dot{\gamma}$ of equal, but otherwise arbitrary norm. This
amounts to deriving the equation of motion from the following
variational principle with constraints

\begin{eqnarray}
\label{eq12} 
\delta\{\;(\;\dot{\gamma}\;
|\;\ln(\gamma\gamma^{+})\;|\;\gamma\;)\;+\;(\;\gamma\;
\;|\;\ln(\gamma\gamma^{+})\;|\;\dot{\gamma}\;)\;+
2\;\zeta\;(\;\dot{\gamma}\;|\;\textbf{H}\;|\;\gamma\;)+%
2\;\zeta^{*}(\;\gamma\;|\;\textbf{H}\;|\;\dot{\gamma}\;)+\nonumber\\
+\;\xi\;[\;(\;\dot{\gamma}\;|\;\gamma\;)+(\;\gamma\;|%
\;\dot{\gamma}\;)\;]+\;\frac{2}{\sigma}\;(\;\dot{\gamma}\;|\;\dot{
\gamma}\;)\;\} = 0\;.
\end{eqnarray}

The variation refers to $\dot{\gamma}$ and $\dot{\gamma}^{+}$ only
and the form of the Lagrange multipliers $\zeta$, $\xi$, $\sigma$
has been chosen for later convenience. $\sigma$ and $\xi$ are real
scalars on account of their corresponding real functionals, while
$\zeta$ is allowed to span complex values. Upon taking the
variation of $\dot{\gamma}$, $\dot{\gamma}^{+}$, one is left with

\begin{equation}
\label{eq13} |\;\dot{\gamma}\;)= -
\;\sigma\;\left[\;\frac{1}{2}\;[\;\ln(\gamma\gamma^{+})\;]\;|\;\gamma\;)
\;+\; \zeta \;\textbf{H}\;|\;\gamma\;)\; +\;
\frac{\xi}{2}\;|\;\gamma\;)\;\right]
\end{equation}

\noindent and the hermitian conjugate. Using Eq.(\ref{eq13}) into
conditions (\ref{eq8a}) and (\ref{eq9a}) immediately gives

\begin{mathletters}
\label{eq14}
\begin{equation}
Re\zeta = -
\frac{1}{2}\;\frac{(\gamma|\;\textbf{H}\;\ln(\gamma\gamma^{+})\;|\gamma)
+ \textrm{E}\;\frac{\textrm{S}}{\textrm{k}_{B}}}{\overline{\Delta
H^{2}}}\;,
\end{equation}

\begin{equation}
\xi  = \frac{\textrm{S}}{\textrm{k}_{B}(\gamma |\gamma)} -
2\;Re\zeta\;\textrm{E}\;,
\end{equation}
\end{mathletters}

\noindent with $\left( {\gamma |\gamma}  \right) =
1\;,\;\textrm{E} = (\gamma|\textbf{H}|\gamma)/(\gamma |\gamma)$
the average energy of the system, $\textrm{S}\geq 0$ the entropy
and $\overline {\Delta H^{2}} = (\gamma |H^{2}|\gamma) -
\textrm{E}^{2}$ the squared energy deviation. One can also check
condition (\ref{eq10a}) and find that

\begin{equation}
\label{eq15} \frac{{\dot {\textrm{S}}}}{{k_{B}} } = \sigma\;
\left( {\theta |\theta}  \right) \;,
\end{equation}

\begin{equation}
\label{eq16}
|\theta ) = \ln (\gamma \gamma ^+)|\;\gamma\; ) +
2\;\zeta\;\textrm{\textbf{H}}|\;\gamma\; ) +
\;\xi\;|\;\gamma\;)\;,
\end{equation}

\noindent
hence inequality (\ref{eq10a}) is satisfied provided

\begin{equation}
\label{eq17}
\sigma \geq  0\;.
\end{equation}

\noindent In deriving expression (\ref{eq15}) we used the fact
that for $|\theta)$ as in Eq.(\ref{eq16}), and $Re\zeta$, $\xi$
given by eqs.(\ref{eq14}), it is also true that

\begin{equation}
\label{eq18}
(\;\gamma\;|\;\textbf{H}\;|\;\theta\;)\;=\; 0\;,\quad
\quad (\;\gamma\;|\;\theta\;) = 0\;.
\end{equation}

\noindent Let us stress at once that, unlike the usual stationary action principle, our variational principle Eq.(\ref{eq12}) does not involve variations of functionals over an extended interval of time, but only variations against $\dot {\gamma}$ which are local in time, at each given instant t. As a result, the Lagrange parameters $\zeta$, $\xi$, $\sigma$ need only be constants against these same variations of $\dot{\gamma}$ and \textit{not} constants of time or $\gamma$ itself. Likewise, condition (\ref{eq17}) for $\sigma$ only guarantees the positivity of the entropy production, but does \textit{not} make $\dot{S}$ independent of time. Hence all parameters in the equation of motion (\ref{eq13}) for $\gamma$, as well as the entropy production and the entropy itself, are time dependent through their dependence on $\gamma$. Furthermore, note that $Re\zeta$, $\xi$ are really functionals of $\rho $ and \textit{H} only and therefore are invariant under transformations of the kind

\begin{equation}
\label{eq19}
\gamma \to \gamma U\;,\quad \quad \quad UU^{+} =
U^{+}U = I\;,
\end{equation}

\noindent
which leave the density matrix unchanged,

\begin{equation}
\label{eq20}
\rho \longrightarrow \rho=\;\gamma
U\;U^{+}\gamma^{+}\;= \gamma \gamma^{+}\;.
\end{equation}

\noindent Eq.(\ref{eq13}) will be invariant in its
entirety under transformations (\ref{eq20}) provided $\sigma$ and $Im\zeta$ are
likewise invariant as  functionals of $\rho$ and \textit{H}. In that case the entropy production Eq.(\ref{eq15}) will also be invariant under transformations (\ref{eq20}), as should be expected on physical grounds.

Now let us introduce the equivalent equation of motion for the
density matrix, starting from

\begin{equation}
\label{eq21}
\dot{\rho}\;=\;\dot{\gamma}\gamma^{+}\;+\;\gamma
\dot{\gamma}^{+} \;.
\end{equation}

\noindent
It follows at once that

\begin{equation}
\label{eq22}
\dot{\rho}= - \sigma\;\left[\;\rho\;\ln\rho\; +
\;Re\zeta\;\{H-\textrm{E},\;\rho\}\;-\;\rho\;Tr(\rho\ln\rho)\right]\;+%
\;i\;\sigma(Im\zeta)[\;\rho,\;H]\;,
\end{equation}

\noindent where \{,\} denotes the anticommutator, as usual. The
commutator on the right hand side of Eq.(\ref{eq22}) obviously
provides the unitary hamiltonian limit, and the standard
Liouville equation suggests

\begin{equation}
\label{eq23}
\sigma(Im\zeta) = \frac{1}{\hbar}\;.
\end{equation}

\noindent Setting now, for simplicity, $Re\zeta \rightarrow \zeta
$, the final form of our equation of motion for the density matrix
is found to be, in common notation,

\begin{equation}
\label{eq24}
\dot{\rho}=-\sigma\left[\;\rho\ln\rho +%
\zeta\;(\rho ,H - \textrm{E})\{H - \textrm{E},\;\rho\}-%
\rho\;\frac{Tr(\rho\ln\rho)}{Tr(\rho)}\right]+\frac{i}{\hbar}\;[\rho,\;H]\;,
\end{equation}

\noindent
where

\[\zeta(\rho,H -\textrm{E})=-\frac{1}{2}\;\frac{Tr[(H -\rm{E})\rho\ln\rho]}%
{Tr[(H -\rm{E})^2\rho]}\;,\]
\[\sigma(\rho,H -\textrm{E})\geq 0\;,\]
\[Tr(\rho)= const.(=1)\;,\]
\[\textrm{E}=\frac{Tr(H\rho)}{Tr(\rho)}=const.\;,\]
\[\dot{\textrm{S}}=-\textrm{k}_{B}\frac{d}{dt}Tr(\rho\ln\rho)\geq
0\;.\]

\noindent
The scale setting parameter $\sigma$ remains unspecified
so far, and will be regarded in the following as a functional of
$\rho$ and \textit{H}. In order to secure that Eq.(\ref{eq24}) is
invariant under a scaling $\rho \rightarrow a\rho $, it must be
assumed that $\sigma(a\rho,H)=\sigma(\rho ,H)$, in which case
scaling invariance is verified straightforwardly. Moreover, since
eq.(\ref{eq24}) should not show a dependence on the zero-point of
the energy, it may be assumed also, as above, that $\sigma =
\sigma(\rho,H-\textrm{E})$. For simplicity, it will be understood
throughout the following that $Tr(\rho)=1$.

It is interesting to note that Eq.(\ref{eq24}) can be recovered
from a modified form of the nonlinear ansatz proposed in Ref.[5],

\[\dot{\rho}=\frac{i}{\hbar}[\rho ,\;H] -
\frac{a}{T}\left[f(\rho)-\rho\frac{Tr(f(\rho))}{Tr(\rho)}\right]\;,\]

\noindent
with the obvious substitutions

\[\frac{a}{T}\rightarrow\sigma$, $f(\rho)\rightarrow\rho\ln\rho +
\zeta\;\{H,\;\rho\}\;.\]

\section{FUNDAMENTAL PROPERTIES OF THE NONLINEAR EVOLUTION}
\label{sec3}

Eq.(\ref{eq24}) secures the hermiticity and positivity of the
density matrix by construction, since it has been generated from
an equation for the state operator \textit{$\gamma $}.
Conversely, Eq.(\ref{eq24}) can be easily decomposed into the
corresponding equations for $\gamma$ and $\gamma^{+}$ by using
the substitution $\rho=\gamma\gamma^{+}$, hence the equations of
motion for $\rho$ and $\gamma$ are indeed equivalent.

Assuming again a well-behaved $\sigma$, Eq.(\ref{eq24}) is seen to
be covariant under time-independent unitary transformations,

\[\rho \rightarrow \tilde{\rho} = U^{+}\rho U\;,\quad
H \rightarrow \tilde{H}= U^{+}HU\;,\]

\noindent and, in particular, invariant under the
(time-independent) symmetry group of the hamiltonian, $[U,\;H]=0$.
But an observable \textit{O} which commutes with \textit{H},
$[H,\;O]=0$, is not, in general, an integral of motion. More
details on the problem follow in Sec. 4.

It is convenient to absorb the hamiltonian commutator term by setting, in
analogy to the usual Heisenberg representation,

\begin{equation}
\label{eq25}
\rho(\textrm{t})=\exp\left[-\frac{i}{\hbar}H\textrm{t}\right]\;\bar{\rho}(\textrm{t})%
\exp\left[\frac{i}{\hbar}H\textrm{t}\right]\;.
\end{equation}

\noindent
Upon substituting expression (\ref{eq25}),
eq.(\ref{eq24}) becomes

\begin{equation}
\label{eq26}
\dot{\bar{\rho}} = -
\sigma\left[\;\bar{\rho}\ln\bar{\rho} + \zeta\;\{H -
\textrm{E},\;\bar{\rho}\} -
\bar{\rho}\;Tr(\bar{\rho}\ln\bar{\rho})\right]\;.
\end{equation}

Now note that for $\bar{\rho}$ corresponding to a pure state,
$\bar{\rho }=\bar{\rho}^{2}=|\Psi\rangle\langle\Psi|$, the
entropy operator vanishes together with the coefficient $\zeta$,
i.e. $\bar{\rho}\ln\bar{\rho}\rightarrow 0$,
$\zeta(H-\textrm{E})\rightarrow 0$, such that $\dot{\bar{\rho}}(
\textrm{t}) = 0$ and $\bar{\rho}( \textrm{t})=\bar{\rho
}(0)=|\Psi\rangle\langle\Psi|$, if $\sigma$ is also finite in this
limit. From Eq.(\ref{eq25}) it follows then that a pure state
evolves into a pure state according to the usual hamiltonian law:

\begin{equation}
\label{eq27}
\rho(\textrm{t})=\rho ^{2}(\textrm{t})=\exp\left[-
\frac{i}{\hbar}H\textrm{t}\right]|\Psi\rangle\langle\Psi|
\exp\left[\frac{i}{\hbar}H\textrm{t}\right]\;.
\end{equation}

\noindent
Another situation where the nonlinear evolution reduces
to the hamiltonian law is found for uniform (equiprobable)
distributions $\rho_{unif} $, when the eigenvalues of the density
matrix are all identical. In this case one has the identity $\bar
{\rho}_{unif}\ln\bar{\rho}_{unif}=\bar{\rho}_{unif}Tr(\bar{\rho}_{unif}\ln\bar{\rho}_{unif})$
and $\zeta(\rho_{unif},H-\textrm{E})\rightarrow 0$, wherefrom
$\dot{\bar{\rho}}_{unif}(\textrm{t})=0,\;\bar{\rho}_{unif}(\textrm{t})=\rho_{unif}(0)$
and

\begin{equation}
\label{eq28}
\rho_{unif}(\textrm{t})=\exp\left[-\frac{i}{\hbar}H\textrm{t}\right]\;\rho_{unif}(0)%
\exp\left[\frac{i}{\hbar}H\textrm{t}\right]\;.
\end{equation}

Recall that under unitary propagation the cardinality of the set
of nonzero eigenvalues of the density matrix is preserved in time.
The same holds true if the density matrix evolves according to
Eq.(\ref{eq24}). In order to see this, let $P_{\nu}=
|\phi_{\nu}\rangle\langle\phi_{\nu}|$ be the projector on some
eigenstate of $\bar{\rho}(\textrm{t}),\;\bar{\rho}\cdot
P_{\nu}=\rho_{\nu}P_{\nu}$, where
$\rho_{\nu}=Tr(P_{\nu}\bar{\rho})$ denotes the corresponding
eigenvalue. Since $Tr(\dot{\bar{\rho}}\cdot P_\nu
)=\dot{\rho}_\nu$, multiplying Eq.(\ref{eq26}) by $P_{\nu}$ and
taking the trace yields

\begin{mathletters}
\label{eq29}
\begin{equation}
\dot{\rho}_{\nu}=-\sigma\;[\;\rho_{\nu}\ln\rho_{\nu}\;+\;%
\alpha_{\nu}(\bar{\rho},H)\;\rho_{\nu}]\; ,
\end{equation}
\begin{equation}
\alpha_{\nu}(\bar{\rho},H)=2\;\zeta(\bar{\rho},H)\;Tr[\;P_{\nu}(\textrm{t})%
(H-\textrm{E})\;]\;+\;\frac{\textrm{S}(\textrm{t})}{\textrm{k}_{B}}\;.\nonumber\\
\end{equation}
\end{mathletters}

\noindent
Taking $\rho_{\nu}\ln\rho_{\nu}\rightarrow 0$ for
$\rho_{\nu}=0$ gives $\dot{\rho}_{\nu}=0$ and
$\rho_{\nu}(\textrm{t})=0$, i.e. a zero eigenvalue evolves into a
zero eigenvalue.

As an immediate corollary, density matrices with a finite number
of ``occupied'' state vectors (i.e. a finite number of nonzero
eigenvalues) are necessarily driven towards a stationary state
with a thermal-like distribution on a finite set of energy
eigenstates. Indeed, in this case the entropy, as a functional of
the eigenvalues $\rho_{\nu}$ and under the constraint of
conserved energy and probability, has a finite absolute maximum.
For this reason, and because $\dot{\textrm{S}}(\textrm{t})\geq 0$
at all times, it can only evolve towards a stationary value less
or equal to that maximum. But, as will be shown,
$\dot{S}(\textrm{t})=0$ implies in fact $\dot{\bar{\rho}}=0$ and
$[\bar{\rho},\;H]=0$, and the stationary version of
Eq.(\ref{eq29}a) gives then the thermal-like distribution. Let us
prove now that $\dot{S}(\textrm{t})=0$ implies stationarity. We
begin by making a change of variables, $\rho_{\nu}=e^{-
\eta_{\nu}},\;\eta_{\nu}\geq 0$, such as to write

\begin{equation}
\label{eq30}
\frac{\textrm{S}}{\textrm{k}_{B}}= \sum\limits_{\nu}
{\eta_{\nu}e^{-\eta_{\nu} }}
\end{equation}

\noindent
and

\begin{equation}
\label{eq31} \frac{\dot{\textrm{S}}}{\textrm{k}_{B}} =
\sum\limits_\nu {(
\dot{\eta_{\nu}}-\eta_{\nu}\dot{\eta_{\nu}})\;e^{-\eta_{\nu}}} =
-\sum\limits_\nu{\eta_{\nu}\dot{\eta_{\nu}}\;e^{ -\eta_{\nu}}} \;,
\end{equation}

\noindent
since
$\sum\limits_{\nu}{\dot{\eta}_{\nu}\;e^{-\eta_{\nu}}} =
\sum\limits_{\nu}{\dot{\rho}_{\nu}}= 0$. Also, Eqs.(\ref{eq29})
give

\begin{equation}
\label{eq32}
\dot{\eta}_{\nu}=-\sigma\;[\;\eta_{\nu}-\alpha_{\nu}\;]  \;,
\end{equation}

\noindent
which taken into Eq.(\ref{eq31}) produces

\begin{equation}
\label{eq33}
\frac{\dot{\textrm{S}}}{\textrm{k}_{B}}=\sigma\;\sum\limits_{\nu}%
{\left[\;\eta_{\nu}^{2}-\alpha_{\nu}\eta_{\nu}\right]\;e^{-\eta_{\nu}}}=%
\sigma\;\sum\limits_{\nu}{\left[\alpha_{\nu}\eta_{\nu}-\alpha_{\nu}^{2}\right]%
\;e^{-\eta_{\nu}}+\frac{1}{\sigma}\sum\limits_{\nu}{(\dot{\eta}_{\nu}^{2}})\;%
e^{-\eta_{\nu}}}\;.
\end{equation}

\noindent Further, use of the explicit expression for
$\alpha_{\nu}$, Eq.(\ref{eq29}b), will show that

\[\sum\limits_{\nu}{\left[\alpha_{\nu}\eta_{\nu}-\alpha_{\nu}^{2}\right]\;e^{-\eta_{\nu}}}=%
\sum\limits_{\nu}{\left[2\;\zeta\;\eta_{\nu}Tr(P_{\nu}(H-\rm{E}))\;e^{-\eta_{\nu}}%
+\frac{\textrm{S}}{\textrm{k}_{B}}\eta_{\nu}\;e^{-\eta_{\nu}}\right]}-
\]

\[-\sum\limits_{\nu}{\left[4\;\zeta ^{2}[Tr(P_{\nu}(H -
\textrm{E}))]^{2} \;e^{-\eta_{\nu}}
+4\;\zeta\frac{\textrm{S}}{\textrm{k}_{B}}Tr(P_{\nu}(H -
\textrm{E}))\;e^{-\eta_{\nu}}+
\left(\frac{\textrm{S}}{\textrm{k}_{B}}\right)^{2}e^{-\eta_{\nu}}\right]}=
\]

\[=- 2\;\zeta Tr((H - \textrm{E})\rho\ln\rho)+
\left(\frac{\textrm{S}}{\textrm{k}_{B}}\right)^{2}- 4\;%
\zeta^{2}Tr\left((H - \textrm{E}\right)^{2}\rho)- \]

\begin{equation}
\label{eq34} -4\;\zeta \frac{\textrm{S}}{\textrm{k}_{B}}Tr((H -
\textrm{E})\rho)-\left(\frac{\textrm{S}}{\textrm{k}_{B}}\right)^{2}
= 0\;,
%\;\;\;\;\;\;\;\;\;\;\;\;\;\;\;\;\; \nonumber\\
%- 4\;\zeta \frac{\textrm{S}}{\textrm{k}_{B}}Tr((H -
%\textrm{E})\rho)-\left(\frac{\textrm{S}}{\textrm{k}_{B}}\right)^{2}
%= 0\;, \;\;\;\;\;\;\;\;\;\;\;\;\;\;\;\;\;\;\;\;\;\;\;\;\;\;\;\;\;\;\;\; \nonumber\\
\end{equation}

\noindent where we have used the explicit expression of $\zeta$,
Eq.(\ref{eq14}a). Accounting for Eq.(\ref{eq34}) in
Eq.(\ref{eq33}) shows that

\begin{equation}
\label{eq35}
\frac{\dot{\textrm{S}}}{\textrm{k}_{B}} =
\frac{1}{\sigma}\sum\limits_{\nu}{\dot{\eta}_{\nu}^{2}\;e^{-\eta_{\nu}}}\;,
\end{equation}

\noindent
wherefrom it follows that $\dot{\textrm{S}}=0$ if and
only if $\dot{\eta}_{\nu}=0$ or, equivalently,
$\dot{\rho}_{\nu}=0$. Consider now that the system is evolving in
an asymptotic region where $\dot{\textrm{S}}(\rm{t})\rightarrow
0$ for all t $ > $ 0 . Since necessarily
$\dot{\rho}_{\nu}\rightarrow 0$, $\dot{\bar{\rho}}$ must be
driven by a unitary evolution,
$\bar{\rho}(\rm{t}\geq\rm{t}_{0})=U(\rm{t})\bar{\rho}(\rm{t}_{0})U^{
+}(\rm{t})$. But for $\dot{\rho}_{\nu}\rightarrow 0$,
Eq.(\ref{eq31}) gives $\ln\rho_{\nu}= -\alpha_{\nu}$, which in
turn shows that

\begin{eqnarray}
\label{eq36}
\bar{\rho}\ln\bar{\rho}=-\sum\limits_{\nu}{\rho_{\nu}\alpha_{\nu}P_{\nu}}=%
-2\;\zeta\;\sum\limits_{\nu}{\rho_{\nu} P_{\nu} Tr(P_{\nu}(H-\textrm{E}))}%
-\sum\limits_{\nu}{\rho_{\nu}P_{\nu}\frac{\textrm{S}}{\textrm{k}_{B}}}= \nonumber\\
= - 2\;\zeta\{H_{D}-\textrm{E},\;\bar{\rho}\}-\frac{\textrm{S}}{\textrm{k}_{B}}\bar{\rho}\;, %
%\;\;\;\;\;\;\;\;\;\;\;\;\;\;\;\;\;\;\;\;\;\;\;\;\;\;\; %
\nonumber \\
\end{eqnarray}

\noindent where $H_{D}=\sum\limits_{\nu}{P_{\nu}Tr(P_{\nu}H)}$ is
the diagonal part of \textit{H} in the eigenbasis of $\bar{\rho}$,
$[H_{D},\;\bar{\rho}]=0$. Introducing the above result into
Eq.(\ref{eq26}), one is lead to

\begin{equation}
\label{eq37} \dot{\bar{\rho}} = -
\sigma\;\zeta\{H_{ND},\;\bar{\rho}\}\;,
\end{equation}

\noindent
with $H_{ND}= H -H_{D}$ the non-diagonal part of
\textit{H} relative to $\bar{\rho}$. But Eq.(\ref{eq37}) cannot
generate a unitary evolution unless $H_{ND}=0$, which implies that
stationary entropy over an extended period of time is equivalent
to

\begin{equation}
\label{eq38}
\;H_{D}=H\;,\\
\end{equation}

\noindent hence $[\bar{\rho},H]=0$ and $\dot{\bar{\rho}}=0$. In
other words, the density matrix of the system (see also
Eq.(\ref{eq25})) is stationary and also diagonal over energy
eigenstates. The explicit form of the occupation probability
corresponding to an (occupied) energy state of energy
$\rm{E}_{\nu}$ follows from Eqs.(\ref{eq29}),

\begin{equation}
\label{eq39}
\rho^{eq}_{\nu}
=\exp\left[-2\;\zeta^{eq}(\textrm{E}_{\nu}-\textrm{E})
-\frac{\textrm{S}^{eq}}{\textrm{k}_{B}}\right]\;,
\end{equation}

\noindent
and can be brought to the recognizable thermal form

\begin{equation}
\label{eq40}
 \rho_{\nu}^{eq}=\frac{1}{\rm{Z}}e^{-\beta \rm{E}_{\nu}}\;,
\end{equation}

\noindent with  $\beta = 2\zeta ^{eq}$ and $\textrm{Z}
=-\beta\textrm{E} + (\textrm{S}^{eq}/\textrm{k}_{B})$.
Surprisingly, the parameter $\zeta $ is seen to become at
equilibrium, up to a factor of 2, the reciprocal temperature
$\beta = 1/\textrm{k}_{B}\textrm{T}$. It should be noted,
nevertheless, that according to our initial assumptions
Eq.(\ref{eq40}) applies only to a finite number of energy
eigenstates and therefore does not refer to a canonical
equilibrium distribution. More precisely, the sign of
$\zeta^{eq}$, and of the generalized temperature T, is not
necessarily positive. For instance, let the occupied energy
eigenstates be labeled by $\nu$ in order of their increasing
energy $\textrm{E}_{\nu}$ and let their total number be N. If the
conserved average energy E is such that

\begin{equation}
\label{eq41} \textrm{E} \geq \frac{1}{N}\sum\limits_{\nu=1}^{N}
{\textrm{E}_{\nu}}\;,
\end{equation}

\noindent
a simple calculation will verify that the entropy will
have an (absolute) maximum, corresponding to the equilibrium
state, on a distribution characterized by a negative
$\zeta^{eq}$, hence a ``negative temperature''.

At this point, let us examine more closely the restrictive
assumption of a finite number of non-vanishing eigenvalues for
the density matrix. It can be noted that it has entered the
argument developed above solely by way of the related assumption
of a finite absolute maximum for the entropy, at the given value
E for the average energy. However, there is good reason to assume
that such an absolute maximum exists at least for a large class of
distributions over infinite sets of (orthogonal) state vectors.
If we can extend this ``finite absolute maximum'' conjecture to
all distributions with a finite average energy, it becomes
possible to generalize the results in Eqs.(\ref{eq38}-\ref{eq40})
and state that the nonlinear dynamics described by
Eqs.(\ref{eq24},\ref{eq26}) drives the system towards an
equilibrium state on energy eigenstates, with thermal-like
occupation probabilities. Of course, when the range of occupied
energy eigenvalues extends to infinity, relation (\ref{eq41}) can
no longer be satisfied for any finite E, and the corresponding
temperature can only be positive.

Finally, we wish to clarify the consistency of the present nonlinear dynamics, which follows a path of  \textit{maximal} entropy production, with Prigogine's celebrated principle of \textit{minimum} entropy production. Let us recall that, according to the latter, physical systems evolve towards stationary states which have minimum entropy production compared to slightly displaced neighboring states. Given that the entropy is a convex functional on the state (configuration) space, bounded from above for any finite average energy, this implies that the physical evolution will take the system towards a local maximum of the entropy or at least towards a ridge. Indeed, in a small enough vicinity of a maximum of the entropy, or of a ridge, \textit{any} evolution with positive entropy production will eventually enter a regime where $\dot{S}$ decreases in time until it vanishes in the equilibrium state or is minimized for the stationary states corresponding to a ridge. The variational principle Eq.(\ref{eq12}) only complements this picture by stating that the evolution should follow the \textit{shortest route} to a state of maximum entropy, i.e. the \textit{direction} of the physical path is selected from among all directions satisfying $\dot{S} \ge 0$ by the requirement that the increase in entropy be maximized at each point in time. In this case it can be said that the entropy production evolves towards a minimum of the maximum, to be attained on a local maximum or a ridge of the entropy (hyper)surface in state space.      

\section{THE LINEAR NEAR-EQUILIBRIUM LIMIT}
\label{sec4}

It is natural to anticipate a linear limit for any nonlinear
dynamics evolving sufficiently close to a canonical thermal
equilibrium state, at least in the high-temperature limit. For
the modified equation of motion proposed here, the linearization
process entails essentially the approximation of the entropy
operator $-\bar{\rho}\ln\bar{\rho}$ to first order in
$\Delta(\bar{\rho}-\rho^{eq})$ around the target equilibrium state

\begin{equation}
\label{eq42}
\rho^{eq}=\frac{1}{\textrm{Z}}e^{-\beta
H}\;,\;\;\ln\textrm{Z}=-\beta
\textrm{E}+\frac{\textrm{S}^{eq}}{\textrm{k}_{B}}\;,
\end{equation}

\noindent
with given average energy E and reciprocal temperature
$\beta$. We proceed from the exact expansion

\begin{equation}
\label{eq43} -\ln\bar{\rho}=\sum\limits_{n=1}^{\infty}
{\frac{1}{n}(I-\bar{\rho})^{n}}\;,
\end{equation}

\noindent
which gives for $\bar{\rho}=\rho
^{eq}+\Delta\bar{\rho}$, and in symmetrized form,

\begin{equation}
\label{eq44}
-(\rho^{eq}+\Delta\bar{\rho})\ln(\rho^{eq}+\Delta\bar{\rho})=%
\frac{1}{2}\left\{(\rho^{eq}+\Delta\bar{\rho}),\;%
\sum\limits_{n=1}^{\infty}{\frac{1}{n}(I-\rho^{eq}-\Delta\bar{\rho})^{n}
} \right\}\;.
\end{equation}

\noindent
Separation of the zero- and first-order terms in
$\Delta\bar{\rho}$ yields

\begin{equation}
\label{eq45}
-(\rho^{eq}+\Delta\bar{\rho})\ln(\rho^{eq}+\Delta\bar{\rho})=%
-\rho^{eq}\ln\rho^{eq}-\frac{1}{2}\left\{\Delta\bar{\rho},\;\ln\rho^{eq}\right\}%
+\frac{1}{2}\left\{\rho^{eq},\;\Lambda(\Delta\bar{\rho})\right\}\;,
\end{equation}

\noindent where $\Lambda(\Delta\bar{\rho})$ represents the
collection of all terms first-order in $\Delta\bar{\rho}$ from the
infinite sum on the right hand side of Eq.(\ref{eq44}). In order
to calculate $\Lambda(\Delta\bar{\rho})$, it is convenient to
define the superoperator \textbf{R} and its tilde-conjugate
$\tilde{\textbf{R}}$ by

\begin{mathletters}
\label{eq46}

\begin{equation}
\textbf{R}\Delta\bar{\rho}=(I-\rho^{eq})\Delta\bar{\rho}\;,
\end{equation}

\begin{equation}
\tilde{\textbf{R}}\Delta\bar{\rho}=\Delta\bar{\rho}(I-\rho^{eq})\;,
\end{equation}

\end{mathletters}

\[\;[\textbf{R},\tilde{\textbf{R}}] = 0\;.\]

\noindent
The expression of $\Lambda(\Delta\bar{\rho})$ can be
obtained now in the compact form

\begin{equation}
\label{eq47}
\Lambda(\Delta\bar{\rho})=-\sum\limits_{n=1}^{\infty}{\frac{1}{n}%
\sum\limits_{m=0}^{n-1}{\textbf{R}^{m}\tilde{\textbf{R}}^{n-m-1}}}\Delta\bar{\rho}\;.
\end{equation}

\noindent
But \textbf{R} has a well-defined inverse and the
super-operator sum in the above expression can be rewritten as

\begin{eqnarray}
\label{eq48}
 -\sum\limits_{n=1}^{\infty}{\frac{\tilde
{\textbf{R}}^{n-1}}{\textit{n}}\sum\limits_{m=0}^{n-1}{\left(\textbf{R}\tilde
{\textbf{R}}^{-1}\right)^{m}}}=\left(\textbf{I}-\textbf{R}\tilde{\textbf{R}}^{-1}\right)^{-1}%
\sum\limits_{n=1}^{\infty}{\frac{\tilde{\textbf{R}}^{n-1}}{\textit{n}}%
\left(\textbf{I}-\left(\textbf{R}\tilde{\textbf{R}}^{-1}\right)^{n}\right)}=\nonumber\\
=\left(\tilde{\textbf{R}}-\textbf{R}\right)^{-1}\sum\limits_{n=1}^{\infty}%
{\left(\frac{\tilde{\textbf{R}}^{n}}{\textit{n}}-\frac{\textbf{R}^{n}}{\textit{n}}\right)}=%
\left(\tilde{\textbf{R}}-\textbf{R}\right)^{-1}
\left[\ln\left(\textbf{I}-\textbf{R}\right)-\ln\left(\textbf{I}
-\tilde{\textbf{R}}\right)\right]\;.\nonumber\\
\end{eqnarray}

\noindent
Taking also into account that

\begin{equation}
\label{eq49}
\rho^{eq}\cdot\Delta\bar{\rho}=(\textbf{I}-\textbf{R})\Delta\bar{\rho}\;,%
\;\;\;\Delta\bar{\rho}\cdot\rho^{eq}=(\textbf{I}-\tilde{\textbf{R}})\Delta\bar{\rho}
\end{equation}

\noindent
and

\begin{equation}
\label{eq50}
\ln(\textbf{I} -
\textbf{R})=-\beta\textbf{H}-(\ln\textrm{Z})\;\textbf{I}\;\;,%
\;\;\ln(\textbf{I}-\tilde{\textbf{R}})=-\beta\tilde{\textbf{H}}-(\ln\textrm{Z})\;\textbf{I}\;,
\end{equation}

\noindent
where

\begin{equation}
\label{eq51}
\textbf{H}\Delta\bar{\rho}=H \cdot
\Delta\bar{\rho}\;\;,\;\;\tilde{\textbf{H}}\Delta\bar{\rho}=\Delta\bar{\rho}\cdot
H\;,
\end{equation}

\noindent
we are lead to:

\begin{equation}
\label{eq52}
\frac{1}{2}\left\{\rho^{eq},\Lambda(\Delta\bar{\rho})\right\}=%
-\frac{\beta}{2}\left(\textbf{H}-\tilde {\textbf{H}}\right)\cdot
\coth\left[\;\frac{\beta}{2}\left(\textbf{H}-\tilde{\textbf{H}}\right)\;\right]
\Delta\bar{\rho}\;.
\end{equation}

Returning to Eq.(\ref{eq45}), the first-order in
$\Delta\bar{\rho}$ approximation to the entropy operator reads now

\begin{eqnarray}
\label{53}
- (\rho^{eq}+\Delta\bar{\rho})\ln(\rho^{eq}+\Delta\bar{\rho})=%
-\rho^{eq}\ln\rho^{eq}-\frac{\beta}{2}\{\Delta\bar{\rho},\;H-\textrm{E}\}%
-\frac{\textrm{S}^{eq}}{\textrm{k}_{B}}\cdot\Delta\bar{\rho}-\nonumber\\
-\frac{\beta}{2}\left(\textbf{H}-\tilde{\textbf{H}}\right)\cdot
\coth\left[\;\frac{\beta}{2}\left(\textbf{H}-
\tilde{\textbf{H}}\right)\;\right]\Delta\bar{\rho}\;.
\end{eqnarray}

\noindent
Note that taking the trace in Eq.(53) gives
$\textrm{S}(\textrm{t})\approx \textrm{S}^{eq}$ in this regime.
Similarly, a simple calculation shows that $\zeta\approx
\beta/2$. Assuming also that
$\sigma\approx\sigma^{eq}=const.(\textrm{E},\;\beta)$ and
inserting everything into Eq.(\ref{eq26}) yields the linearized
equation of motion

\begin{equation}
\label{eq54}
\dot{\bar{\rho}}=-\sigma^{eq}\frac{\beta}{2}\left(\textbf{H}-\tilde{\textbf{H}}\right)%
\cdot\coth\left[\;\frac{\beta}{2}\left(\textbf{H}-\tilde{\textbf{H}}\right)\;\right]
\Delta\bar{\rho}
\end{equation}

\noindent
or , as well,

\begin{equation}
\label{eq55}
\Delta\dot{\bar{\rho}}=-\sigma^{eq}\frac{\beta}{2}\left(\textbf{H}-\tilde{\textbf{H}}\right)
\cdot\coth\left[\;\frac{\beta}{2}\left(\textbf{H}-\tilde{\textbf{H}}\right)\;\right]%
\Delta\bar{\rho}\;.
\end{equation}

The general solution of Eq.(\ref{eq54}) is given by

\begin{equation}
\label{eq56}
\bar{\rho}\left(\textrm{t}\right)=\textrm{e}^{-\sigma^{eq}\left(\beta\right)\rm{t}}%
\textrm{e}^{-\bf{G}\rm{t}}\bar{\rho}\left(0\right)+%
\left(1-\textrm{e}^{-\sigma^{eq}\left(\beta\right)\rm{t}}\right)\rho^{eq}\;,
\end{equation}

\noindent
where

\begin{equation}
\label{eq57} \textbf{G} =
\sigma^{eq}(\beta)\frac{\beta}{2}\left(\textbf{H}-\tilde{\textbf{H}}\right)\cdot%
\coth\left[\frac{\beta}{2}\left(\textbf{H}-\tilde{\textbf{H}}\right)\right]-\textbf{I}\;.
\end{equation}

\noindent We observe immediately that \textbf{G} is
tilde-symmetric, hence it maps any hermitian operator into a
hermitian operator, and that it preserves probability, since
$Tr[\textbf{G}\bar{\rho}]=0$,
$Tr[\textrm{e}^{-\bf{G}\rm{t}}\bar{\rho}(0)]=Tr[\bar{\rho}(0)]=1$.
This is entirely sufficient to secure the hermiticity of
$\bar{\rho}$ and the overall conservation of probability.
Unfortunately, the action of \textbf{G} does not always preserve
positivity and \textbf{G} cannot be identified as a generator of
Lindblad type \cite{12}. But the positive domain of \textbf{G}
does include the small neighborhood of $\rho^{eq}$ identified as
the near-equilibrium domain. Indeed, note first that in the
diagonal representation of the hamiltonian, the matrix elements of
$\bar{\rho}^{0}(\textrm{t})=\textrm{e}^{-\bf{G}\rm{t}}\bar{\rho}(0)$
obey the simple damping law

\begin{equation}
\label{eq58}
\bar{\rho}_{\mu\nu}^{0}(\rm{t})=\textrm{e}^{-\gamma_{\mu\nu}%
(\beta)\textrm{t}}\bar{\rho}_{\mu\nu}(0)\;,
\end{equation}

\noindent
where the (temperature-dependent) relaxation
coefficient $\gamma_{\mu\nu}$ is given by

\begin{equation}
\label{eq59}
\gamma_{\mu\nu}(\beta)=\sigma^{eq}(\beta)\left[\frac{\beta}{2}%
\left(\textrm{E}_{\mu}-\textrm{E}_{\nu}\right)\coth\left[\frac{\beta}{2}%
(\textrm{E}_{\mu}-\textrm{E}_{\nu})\right]-1\right]\;,
\end{equation}

\[\gamma_{\nu\nu}=0\;,\;\gamma_{\nu\mu}=\gamma_{\mu\nu}\;.\]

\noindent If we consider now an arbitrary state vector
$|\Psi\rangle=\sum\limits_{\nu=0}^{\infty} {\langle \textrm{E}
_{\nu}|\Psi\rangle|\textrm{E}_{\nu}\rangle}$ and the matrix
element

\begin{equation}
\label{eq60}
\left\langle\Psi|\bar{\rho}^{0}(\textrm{t})|\Psi\right\rangle =
\sum\limits_{\nu=0}^{\infty}{\left\langle\Psi
|\textrm{E}_{\nu}\right\rangle
\bar{\rho}_{\nu\nu}(0)\left\langle\textrm{E}_{\nu}|\Psi\right\rangle}
+ \sum\limits_{\mu,\nu = 0\hfill\atop\mu > \nu
\hfill}^{\infty}{Re\left[\left\langle\Psi|\textrm{E}_{\mu}\right\rangle
\bar{\rho}_{\mu\nu}(0)\left\langle\textrm{E}_{\nu}|\Psi\right\rangle
\right]\;\textrm{e}^{-\gamma_{\mu\nu}\rm{t}}}\;.
\end{equation}

\noindent it is easily seen that $\bar{\rho}(\textrm{t})$ remains
positive for $\textrm{t}>0$ if the initial off-diagonal matrix
elements $\bar{\rho}_{\mu\nu}(0)$, $\mu \neq \nu$, are
sufficiently small, as expected for the near-equilibrium regime.
On the other hand, one can resort to the equation of motion for
the state operator $\gamma$, Eq(\ref{eq13}), and derive a linear
approximation in $\Delta\gamma=\gamma-\gamma^{eq}$,
$\Delta\gamma^{+}=\gamma^{+}-(\gamma^{eq})^{+}$ by the same
procedure as above. The result reads

\begin{mathletters}

\begin{equation}
\dot{\Delta\gamma}=-\left[\left(\sigma \beta + \frac{i}{\hbar }\right)%
\left(\textbf{H}-\tilde{\textbf{H}}\right)\left(\tilde{\textbf{R}}%
-\textbf{R}\right)^{-1}\left(\gamma^{eq}\Delta\gamma^{+}%
+\Delta\gamma\left(\gamma^{eq}\right)^{+}\right)\right]\gamma^{eq}
\end{equation}

\begin{equation}
\dot{\Delta\gamma}^{+}=-\left(\sigma\beta - \frac{i}{\hbar }\right)%
\left(\gamma^{eq}\right)^{+}\left[\left(\textbf{H}-\tilde{\textbf{H}}\right)\left(\tilde{\textbf{R}}%
-\textbf{R}\right)^{-1}\left(\gamma^{eq}\Delta\gamma^{+}%
+\Delta\gamma\left(\gamma^{eq}\right)^{+}\right)\right]
\end{equation}

\end{mathletters}

\noindent and shows that, up to first-order terms in
$\Delta\gamma$,

\begin{equation}
\rho(\textrm{t})=\rho^{eq}+ \Delta\rho \approx \left[\gamma^{eq}
+\Delta\gamma\right]\left[\left(\gamma^{eq}\right)^{+} +
\Delta\gamma^{+}\right]
\end{equation}

\noindent such that $\Delta\rho =
\Delta\gamma\left(\gamma^{eq}\right)^{+} +
\gamma^{eq}\Delta\gamma^{+}$ evolves according to Eq.(\ref{eq55})
derived above. Furthermore, the conservation of energy follows
from

\begin{equation}
\label{eq63}
Tr\left(H\Delta\dot{\bar{\rho}}\right)=-\sigma
^{eq}Tr\left(H\Delta\bar{\rho}\right)
\end{equation}

\noindent upon recalling that, according to the original equation
of motion, the initial state necessarily has the same average
energy E as the asymptotic equilibrium state. The initial
conditions for Eq.(\ref{eq56}) are so restricted to
$Tr\left(H\Delta\bar{\rho}(0)\right)=0$, which implies of course
$Tr\left(H\bar{\rho}(0)\right)=\textrm{E}$.

As a general feature of the underlying physics, it follows from
Eqs.(\ref{eq56}), (\ref{eq58}) and (\ref{eq59}) that the greater
the energy gap between two energy eigenstates, the faster the
quantum correlation between them is destroyed as the system
evolves towards equilibrium. On the other hand, the relaxation of
the occupation probabilities for each of the energy states
proceeds at a common rate, independent of the corresponding
energy level, since
$\bar{\rho}_{\nu\nu}(\rm{t})=\frac{\textrm{e}^{-\beta\rm{E}_{\nu}}}{\textrm{Z}}
\left(1-\textrm{e}^{-\sigma^{eq}\left(\beta\right)\rm{t}}\right)
+ \textrm{e}^{-\sigma^{eq}\left(\beta\right)\rm{t}}
\bar{\rho}_{\nu\nu}(0)$. As a corrolary, the same holds true for
the average of any observable \textit{O} which commutes with the
hamiltonian, $\left[H,\;O\right] = 0$, since $\langle
O(\textrm{t})\rangle = Tr[O\rho(\textrm{t})] =
Tr\left[O\;\textrm{e}^{-\frac{i}{\hbar}H\rm{t}}\bar{\rho}(\textrm{t})
\;e^{\frac{i}{\hbar}H\rm{t}}\right]$ (see Eq.(\ref{eq26})) will
involve only $\bar{\rho}_{\nu \nu}$-s. The same result can be
obtained in a formal manner from a generalized Heisenberg
representation for Eq.(\ref{eq54}), in which the observables
evolve in time according to

\begin{mathletters}
\label{eq64}

\begin{equation}
\dot{O}_{\Delta}(\textrm{t})=-\left[\sigma^{eq}(\beta)\left(\textbf{G}%
\left(\frac{\beta}{2}\left(\textbf{H}-\tilde{\textbf{H}}\right)\right)+ \textbf{I} \right)%
-\frac{i}{\hbar}\left(\textbf{H}-\tilde{\textbf{H}}\right)\right]\;O_{\Delta}(\textrm{t})%
\end{equation}

\begin{equation}
O_{\Delta}(\textrm{t})=\exp\left\{-\left[\sigma^{eq}(\beta)%
\left(\textbf{G}\left(\frac{\beta}{2}\left(\textbf{H}-\tilde{\textbf{H}}\right)\right)%
+ \textbf{I}\right)-\frac{i}{\hbar }\left(\textbf{H}-\tilde{\textbf{H}}\right)\right]%
\;\textrm{t}\right\}\;O_{\Delta}(0)\;.
\end{equation}

\end{mathletters}

\noindent
Here $\left(\textbf{H}-\tilde{\textbf{H}}\right)O =
[H,\;O]$ and the lower label $\Delta$ reminds that all averages
are to be calculated with $\Delta\rho(0)=\rho(0)-\rho^{eq}$. From
Eq.(\ref{eq64}) above it is immediate that
$[H,\;O]=\left(\textbf{H}-\tilde{\textbf{H}}\right)O = 0$ yields

\begin{mathletters}
\label{eq65}

\begin{equation}
\dot{O}_{\Delta}(\textrm{t})=-\sigma^{eq}O_{\Delta}(\textrm{t})\;,
\end{equation}

\begin{equation}
\;\;O_{\Delta}(\textrm{t})=
\exp\left[-\sigma^{eq}\textrm{t}\right]\;O_{\Delta}(0)\;,
\end{equation}

\end{mathletters}

\noindent in agreement with the observation above. An unexpected
outcome of this result is that the average of an observable which
commutes with the hamiltonian is conserved throughout the
evolution, provided the initial average value is identical to the
equilibrium average. In other words, $\langle O_{Delta}(0)\rangle$
implies $\langle O \rangle (\textrm{t})= \langle O \rangle^{eq}$.
The conservation of energy, Eq.(\ref{eq63}), is seen to be in fact
just a particular realization of this feature. Furthermore, for
operators satisfying commutation relations of the form
$\left[H,\;A\right]=\varepsilon A$, Eqs.(\ref{eq64}) lead to

\begin{mathletters}
\label{eq66}

\begin{equation}
 \dot {A}_{\Delta}  \left( {t} \right) = - \left[ {\sigma ^{eq}\left( {\beta
} \right)\left( {G\left( {\frac{{\beta \varepsilon} }{{2}}}
\right) + 1} \right) - \frac{{i}}{{\hbar} }\varepsilon}
\right]A_{\Delta}  \left( {t} \right)\;,
\end{equation}

\begin{equation}
 A_{\Delta}  \left( {t} \right) = \exp\left[ { - \left[ {\sigma ^{eq}\left(
{\beta}  \right)\left( {G\left( {\frac{{\beta \varepsilon}
}{{2}}} \right) + 1} \right) - \frac{{i}}{{\hbar} }\varepsilon}
\right]t} \right]A_{\Delta} \left( {0} \right)\;,
\end{equation}

\end{mathletters}

\noindent
where $\textbf{G}(x)+1=x\;\coth(x)$.

Eqs.(\ref{eq65}) and (\ref{eq66}) allow us to provide a handful of
instant examples.

1) A two-level atom, with the hamiltonian

\[H=\textrm{E}_1|1\rangle\langle 1|+\textrm{E}_2|2\rangle\langle
2|\]

\noindent
and the occupation numbers

\[n_{1}=\langle1|\rho|1\rangle\; , \;
n_{2}=\langle2|\rho|2\rangle\;,\;\; n_{1}+n_{2}=1\;,\]

\noindent obeys a simple relaxation law which follows from
Eq.(\ref{eq65}a):

\begin{mathletters}
\label{eq67}

\begin{equation}
\dot{n}_{1}=-\sigma^{eq}(\beta)\left(n_{1}-n_{1}^{eq}(\beta)\right)\\
\end{equation}

\begin{equation}
\dot{n}_{2}=-\sigma^{eq}(\beta)\left(n_{2}-n_{2}^{eq}(\beta)\right)\\
\end{equation}

\end{mathletters}

%\[n_{1}+n_{2}=n_{1}^{eq}+n_{2}^{eq}=1.\]

\noindent
If Eqs.(\ref{eq67}) are rearranged into the kinetic-like
form

\begin{mathletters}
\label{eq68}

\begin{equation}
\dot{\textrm{n}}_1 = -\textrm{k}_{12}\;\textrm{n}_1 +
\textrm{k}_{21}\;\textrm{n}_2\; ,
\end{equation}

\begin{equation}
\dot{\textrm{n}}_2 =  \;\textrm{k}_{12}\;\textrm{n}_1 -
\textrm{k}_{21}\;\textrm{n}_2 \; ,
\end{equation}

\end{mathletters}

\noindent
the corresponding (thermal) transition rates
$\textrm{k}_{12}=\sigma^{eq}(\beta)\textrm{n}_{2}^{eq}(\beta)
\;,\;\textrm{k}_{21}=\sigma^{eq}(\beta)\textrm{n}_{1}^{eq}(\beta)$,
are seen to have, up to the factor of $\sigma^{eq}$, an
Arrhenius-like dependence on the temperature.

2) For a harmonic oscillator of unit mass and frequency $\omega$,
described by

\[ H = \frac{{p}^{2}}{{2}} + \frac{\omega^{2}q^{2}}{2}\;,\]
\[\left\langle p\right\rangle_{eq}=0,\;\left\langle
q\right\rangle_{eq}=0 \;,\]

\noindent one can apply Eq.(\ref{eq66}a) to the annihilation and
creation operators,

\[a=\sqrt{\frac{\omega}{2\;\hbar}}\left(q+i\frac{p}{\omega}\right)\;,\;\;
a^{+} = \sqrt{\frac{\omega}{2\;\hbar}}\left(q
-i\frac{p}{\omega}\right)\;,\]

\noindent
to recover a coupled system of equations for the average
momentum and the average coordinate,

\begin{mathletters}
\label{eq69}

\begin{equation}
\left\langle\dot{p}\right\rangle =- \gamma(\omega,\;\beta)\langle p \rangle%
 - \omega^{2}\langle q \rangle\; ,
\end{equation}

\begin{equation}
\left\langle\dot{q}\right\rangle = \langle p \rangle - %
\gamma(\omega,\;\beta)\langle q \rangle  \; ,
\end{equation}

\end{mathletters}

\noindent
where
$\gamma\left(\omega,\beta\right)=\sigma^{eq}\left(\beta\right)\left[1
+ G\left(\beta\hbar\omega/2\right)\right]$. We recognize a
typical damped motion, driven by the classical Langevin equation

\begin{equation}
\label{eq70} \left\langle {\ddot {q}} \right\rangle + 2\gamma
\left( {\omega ,\beta} \right)\left\langle {\dot {q}}
\right\rangle + \left[ {\omega ^{2} + \gamma ^{2}\left( {\omega
,\beta}  \right)} \right]\left\langle {q} \right\rangle = 0 \;,
\end{equation}

\noindent
which is obtained by elimination of the momentum
variables from Eqs.(\ref{eq69}).

3) For the non-relativistic free particle hamiltonian

\[H = \frac{p^{2}}{2m}\]

\noindent Eq.(\ref{eq65}a) gives the relaxation law

\begin{equation}
\label{eq71} \left\langle {\dot {p}} \right\rangle = - \sigma
^{eq}\left[ {\left\langle {p} \right\rangle - \left\langle {p}
\right\rangle ^{eq}} \right] \;,
\end{equation}

\noindent
which shows a (thermal) friction force linear in momentum. In case the
initial momentum average coincides with the final thermal average, one
obviously obtains conservation of the average momentum. More details can be
extracted from the Wigner function

\begin{equation}
\label{eq72} f_\Delta  ({\rm \vec p},{\rm \vec r},{\rm t}) =
\sum\limits_{{\rm \vec q}} {e^{ - \frac{i}{\hbar }{\rm \vec q}
\cdot {\rm \vec r}} \left\langle {{\rm \vec p} - \frac{{{\rm \vec
q}}}{{\rm 2}}} \right|\rho _\Delta  } ({\rm t})\left| {{\rm \vec
p + }\frac{{{\rm \vec q}}}{{\rm 2}}} \right\rangle\;.
\end{equation}

\noindent
Differentiation of Eq.(\ref{eq72}) on time and use of
Eqs.(\ref{eq56}), (\ref{eq58}), (\ref{eq59}) yields

\begin{equation}
\label{eq73} \dot{f}_{\Delta}\left( {\vec {p},\vec {r},t} \right)
= \sum\limits_{\vec {q}} {e^{ - \frac{{i}}{{\hbar} }\vec {q}
\cdot \vec {r}}\left[ { - \sigma ^{eq}\left( {\beta}
\right)\left( {G\left( {\beta \frac{{\vec {p} \cdot \vec
{q}}}{{m}}} \right) + 1} \right) + \frac{{i}}{{\hbar}
}\frac{{\vec {p} \cdot \vec {q}}}{{m}}} \right]\left\langle {\vec
{p} - \frac{{\vec {q}}}{{2}}} \right|\rho _{\Delta}  \left( {t}
\right)\left| {\vec {p} + \frac{{\vec {q}}}{{2}}}
\right\rangle}\;.
\end{equation}

\noindent
But note that

\begin{equation}
\label{eq74} \frac{{\vec {p}}}{{m}} \cdot \nabla _{\vec {r}}
f_{\Delta}  \left( {\vec {p},\vec {r},t} \right) =
\sum\limits_{\vec {q}} {e^{ - \frac{{i}}{{\hbar }}\vec {q} \cdot
\vec {r}}\left( { - \frac{{i}}{{\hbar} }\frac{{\vec {p} \cdot
\vec {q}}}{{m}}} \right)\left\langle {\vec {p} - \frac{{\vec
{q}}}{{2}}} \right|\rho _{\Delta}  \left( {t} \right)\left| {\vec
{p} + \frac{{\vec {q}}}{{2}}} \right\rangle}
\end{equation}

\noindent
and rewrite the right hand side of Eq.(\ref{eq73}) in
differential form to obtain

\begin{equation}
\label{eq75} \dot {f}_{\Delta}  + \frac{{\vec {p}}}{{m}} \cdot
\nabla _{\vec {r}} f_{\Delta}  = - \sigma ^{eq}\left( {\beta}
\right)\left( {G\left( {i\frac{{\hbar \beta} }{{m}}\vec {p} \cdot
\nabla _{\vec {r}}}  \right) + 1} \right)f_{\Delta}\;.
\end{equation}

\noindent
The operatorial expression on the right hand side is to
be understood in terms of the power expansion $G\left( {x}
\right) + 1= x\;\coth\left( {x} \right) = 1 + 2\sum\limits_{n =
1}^{\infty} {\left( { - 1} \right)^{n - 1}\zeta _{R} \left( {2n}
\right)( x/\pi)^{2n}} $ \cite{13}, where $\zeta _R ({\rm s}) =
\sum\limits_{{\rm k} = 1}^\infty {{\rm k}^{{\rm  - s}} } $ is the
Riemann zeta function. Hence Eq.(\ref{eq75}) reads, in explicit
form,

\begin{equation}
\label{eq76}
\dot {f}_{\Delta}  + \frac{{\vec {p}}}{{m}} \cdot
\nabla _{\vec {r}} f_{\Delta}  = \sigma ^{eq}\left( {\beta}
\right)\left[ { - 1 + 2\sum\limits_{n = 1}^{\infty}  {\zeta _{R}
\left( {2n} \right)\left( {\frac{{\hbar \beta} }{{\pi m}}\vec {p}
\cdot \nabla _{\vec {r}}} \right)^{2n}}}  \right]f_{\Delta}
\end{equation}

\noindent
and proves to be a Burnett-type (or generalized
Fokker-Plank) equation. Note further that the dependence of
$f_{\Delta} $ on momentum and coordinate variables can be
separated in Eq.(\ref{eq76}), and one can integrate over momentum
to obtain an exact equation in coordinate space. It can be safely
assumed also that the momentum distribution does not deviate
significantly from equilibrium, such that one can write
$f_{\Delta}  \left( {\vec {p},\vec {r},t} \right) \approx
n_{\Delta}  \left( {\vec {r},t} \right)f^{eq}\left( {\left| {\vec
{p}} \right|} \right)$, where $n_{\Delta}  \left( {\vec {r},t}
\right) = n\left( {\vec {r},t} \right) - n^{eq}\left( {\vec
{r},t} \right)$ is the deviation from the equilibrium value of
the localization probability and $f^{eq}\left( {\left| {\vec {p}}
\right|} \right)$ is the equilibrium momentum distribution. In
that case, in the high-temperature limit when only contributions
to leading order in \textit{$\beta $} survive, integration over
momentum leads apparently to a diffusion-like equation,

\begin{equation}
\label{eq77}
{\rm \dot n}_\Delta   = {\rm D}\left( \beta
\right)\Delta {\rm n}_\Delta   - \sigma ^{eq} (\beta ){\rm
n}_\Delta\;,
\end{equation}

\noindent
with the diffusion coefficient

\begin{equation}
\label{eq78}
D\left( {\beta}  \right) = \sigma ^{eq}\left(
{\beta}  \right)\frac{{\hbar ^{2}\beta} }{{3m}} \; ,
\end{equation}

\noindent
where it has been taken into account that $\zeta_{R}(2)
= \pi^{2}/6$. But let us recall that $\lambda _{T} = \sqrt {(\hbar
^{2}\beta)/(3\;m)} $ is just the de Broglie wavelength
corresponding to the root-mean-square momentum $\sqrt
{\left\langle \vec{p}^{2} \right\rangle^{eq}} $, such that in
fact $\textrm D(\beta) = \sigma
^{eq}(\beta)\left(\lambda_T\right)^2 $. It follows necessarily
that the diffusion term in Eq.(\ref{eq77}) can give significant
contributions only if the localization probability varies
substantially on the scale of the thermal de Broglie wavelength
\textit{$\lambda $}$_{T}$, regardless of the specific value of
$\sigma ^{eq}$(\textit{$\beta $}). But since states with such
variations do not belong to the high-temperature,
near-equilibrium regime, we are forced to recognize that
Eq.(\ref{eq77}) actually reduces to

\begin{equation}
\label{eq79}
\dot {n}_{\Delta}  = - \sigma ^{eq}\left( {\beta}
\right)n_{\Delta}\;.
\end{equation}

The linearization procedure developed in this Section can be
extended without significant modifications to equilibrium states
other than the thermal canonical distribution. It can be shown
that the relaxation laws for the elements of the density matrix
in the diagonal representation of the hamiltonian are similar to
those found here for the canonical case. A detailed account of
this issue will be given elsewhere.

\section{SYMMETRY INVARIANCE, CONSERVATION LAWS AND SEPARABILITY}
\label{sec5}

It has already been pointed out in Sec.3 that Eq.(\ref{eq24}) is
invariant under any time-independent unitary transformations that
leave the hamiltonian unchanged. It is also obviously invariant
against time translations, albeit this operation can no longer be
associated with a unitary transformation. The same is not true,
in this form, of time-dependent transformations relating
different observers in relative motion. But at least in the
non-relativistic case, this deficiency can be easily corrected so
that invariance under the complete dynamical group of the system
is recovered. Indeed, let us rewrite Eq.(\ref{eq24}) in the form

\begin{equation}
\label{eq80}
\dot {\rho}  = - \sigma \left[ {\rho ln\rho + \left\{
{D\left( {\rho} \right),\rho}  \right\} - \rho\; Tr\left( {\rho
ln\rho}  \right)} \right] + \frac{{i}}{{\hbar} }\left[ {\rho ,H}
\right]\;,
\end{equation}

\noindent
where $D(\rho)$ replaces $\zeta(\rho,\;H -
\textrm{E})\;(H - \textrm{E})$, and let us consider the invariance
conditions for Eq.(\ref{eq80}) under a time-dependent unitary
transformation $U(\textrm{t})$, $U\left( {t} \right)U^{ +} \left(
{t} \right) = U^{ +} \left( {t} \right)U\left( {t} \right) = I$.
As usual, the density matrix becomes ${\rho }'\left( {t} \right)
= U\left( {t} \right)\rho \left( {t} \right)U^{ + }\left( {t}
\right)$, hence

\begin{equation}
\label{eq81}
\dot{\rho}'(\textrm{t}) =
U(\textrm{t})\dot{\rho}(\textrm{t})U^{+}(\textrm{t})-\left[
\rho'(\textrm{t}),\;\dot{U}(\textrm{t})U^{+}(\textrm{t})\right]\;
,
\end{equation}

\noindent
while multiplication of Eq.(\ref{eq80}) by
$U(\textrm{t})$ on the left and $U^{+}(\textrm{t})$ on the right,
followed by use of Eq.(\ref{eq81}) gives

\begin{eqnarray}
\label{eq82}
\dot{\rho}'
=-\sigma\left[\rho'\ln\rho'+\left\{U(\textrm{t})D(\rho)U^{+}(\textrm{t}),\;\rho'\right\}%
- \rho'\;Tr\left(\rho'\ln\rho'\right)\right] + \nonumber\\
+\frac{i}{\hbar}\left[\rho',\;U(\textrm{t})HU^{+}(\textrm{t}) +
i\hbar\;\dot{U}(\textrm{t})U^{+}(\textrm{t})\right]\;.
\end{eqnarray}

\noindent
It is easily seen that Eq.(\ref{eq82}) will regain the
form of Eq.(\ref{eq80}) provided \textit{H} is invariant under
$U(\textrm{t})$ in the customary sense,

\begin{equation}
\label{eq83}
U(\textrm{t})HU^{+}(\textrm{t}) +
i\hbar\;\dot{U}(\textrm{t})U^{+}(\textrm{t}) = H
\end{equation}

\noindent
and if, in addition,

\begin{mathletters}
\label{eq84}

\begin{equation}
U(\textrm{t})D(\rho)U^{+}(\textrm{t})=D(\rho')\; ,
\end{equation}

\begin{equation}
\sigma(\rho ,\;H - \textrm{E}) = \sigma (\rho',\;H -
\textrm{E}')\; ,
\end{equation}

\end{mathletters}

\noindent
In the absence of any evidence to the contrary, the
functional $\sigma$ will be assumed in the following to have all
necessary invariance properties.

From Eq.(\ref{eq83}) it follows in the customary way that if
$U(\textrm{t})$ spans a Lie group of order n, such that
$U(\textrm{t})=\exp\left[(i/\hbar)\lambda^{j}K_{j}(\textrm{t})
\right]$, with $\lambda^{j},\; j = 1,\;2,\ldots n$ the group
parameters and summation over repeated indices understood, then
the corresponding infinitesimal, hermitian generators
$K_{j}(\textrm{t}),\;j = 1,\;2,\ldots n$ satisfy the familiar
commutation relations

\begin{equation}
\label{eq85}
\left[ {K_{j} \left( {t} \right),H} \right] -
\frac{{\partial} }{{\partial t}}K_{j} \left( {t} \right) = 0\;.
\end{equation}

\noindent
Note that a conservation law is not yet implied. But let
us assume further that the transformations $U(\textrm{t})$ are
such that

\begin{mathletters}
\label{eq86}

\begin{equation}
\dot{U}(\textrm{t})U^{+}(\textrm{t}) = \frac{i}{\hbar}\left(\;a^{j}C_{j}(\textrm{t})%
+ b\; \right)\; ,
\end{equation}

\begin{equation}
U(\textrm{t})C_{j}(\textrm{t})U^{+}(\textrm{t}) = c_{j}^{\;\;l}
\;C_{l}(\textrm{t}) + f_{j}\; ,
\end{equation}

\end{mathletters}

\noindent
where all parameters $a^{j},\;b,\;c_{j}^{\;\;l} ,f_{j} $
are real functions of the group parameters $\lambda ^{j}$ and
time, and the $C_{j}$-s are hermitian operators (observables). In
that case, if the conservation of energy is to be invariant under
all transformations $U(\textrm{t})$, it follows from the
expression of the transformed average energy

\begin{equation}
\label{eq87}
{\textrm{E}}' = Tr\left( {H{\rho} '\left( {t}
\right)} \right) = \textrm{E} + i\hbar \;Tr\left( {\dot {U}\left(
{t} \right)U^{ +} \left( {t} \right){\rho}'\left( {t} \right)}
\right) = \textrm{E} - a^{j}\left\langle {C_{j}}\right\rangle
^{\prime}  - b \;,
\end{equation}

\noindent
that a conservation law is required for each $C_{j}$.
Unfortunately, Eq.(\ref{eq24}) does not account for such
supplementary constants of motion and simple algebra reveals that
$D(\rho)=\zeta(\rho,\;H - \textrm{E})\;(H - \textrm{E})$ does not
satisfy the first of Eqs.(\ref{eq84}), despite an invariant
hamiltonian, since

\[ U\left( {t} \right)\left[ {\zeta \left( {\rho ,H - \textrm{E}}
\right)\; \left( {H - \textrm{E}} \right)} \right]U^{ +} \left(
{t} \right) = \]

\[ = \zeta \left( {{\rho }',H - i\hbar \;\dot
{U}\left( {t} \right)U^{ +} \left( {t} \right) - \textrm{E}}
\right)\left( {H - i\hbar \;\dot {U}\left( {t} \right)U^{ +}
\left( {t}\right) - \textrm{E}} \right) \]

\begin{equation}
\label{eq88}
\ne \zeta \left( {{\rho} ',H - \textrm{E}'}
\right)\left( {H - \textrm{E}'} \right)\;.
\end{equation}

Let us examine now whether modifying Eq.(\ref{eq80}) to include
conservation of the quantities \textit{C}$_{j}$ brings about the
desired invariance under the transformations of the given Lie
group. Let the conservation of each \textit{C}$_{j}$ be added to
the set of constraints accounted for in the original variational
principle, such that Eq.(\ref{eq12}) is brought to the form

\begin{eqnarray}
\label{eq89}
 \delta \left[ \;(\;\dot \gamma \;|\;\ln (\gamma \gamma ^ +  )\;|\;\gamma\; ) %
 + (\;\gamma\; |\;\ln (\gamma \gamma ^ +  )\;|\;\dot \gamma \;) +
  2\;\zeta\; (\;\dot \gamma\; |\;{\bf H}\;|\;\gamma\; ) + %
 2\;\zeta^{*}(\;\gamma\; |\;{\bf H}\;|\;\dot \gamma\; )  +\right.\nonumber\\
\\
\left. + \;\xi\; \left(\; (\;\dot \gamma \;|\;\gamma\; )+
(\;\gamma\; |\;\dot \gamma\; )\;\right)
    + 2\;\eta ^j \left(\; (\;\dot \gamma \;|{\bf C}_j\; |\;\gamma\; )%
   + (\;\gamma\; |\;{\bf C}_j \;|\dot \gamma \;) \;\right) + \frac{2}{\sigma }\;%
   (\;\dot \gamma \;|\;\dot \gamma\; )\right] = 0\;, \nonumber
\end{eqnarray}

\noindent
with the new parameters $\eta^{j}$ assumed real, since
the corresponding terms will not contribute to the hamiltonian
part of the equation of motion. Taking again the variation with
respect to $\dot{\gamma},\;\dot{\gamma}^{+}$ yields

\begin{equation}
\label{eq90} |\dot {\gamma}) = - \sigma \left[ \frac{1}{2}\left[
\ln\left(\gamma \gamma ^{ +}\right)\right]|\gamma) + \zeta\;
 \textbf{H}|\gamma) + \eta^{j}\textbf{C}_{j}\;|\gamma)
+ \frac{\xi}{2}\;|\gamma) \right]
\end{equation}

\noindent
and the corresponding equation of motion for the density matrix,

\begin{equation}
\label{eq91} \dot{\rho}=- \sigma \left[\rho\ln\rho
+\left\{\zeta\;(H-\textrm{E})+\eta^{j}\left(C_{j}-\left\langle
C_{j}\right\rangle\right),\;\rho\right\} -
\rho\;Tr(\rho\ln\rho)\right]+\frac{i}{\hbar}[\rho ,\;H]\;.
\end{equation}

\noindent
Here $\left\langle C_{j}\right\rangle =
Tr\left[C_{j}\rho\right]$ is the conserved average of $C_{j}$,
$\zeta $ and $\eta^{j}$ are solutions of

\begin{mathletters}
\label{eq92}

\begin{equation}
 Tr\left[(H - \textrm{E})\rho\ln\rho\right]+ 2\zeta \;Tr\left[\left(H -
\textrm{E}\right)^{2}\rho\right]+\eta^{j}\;Tr\left[\left\{H-\textrm{E},\;C_{j}
-\left\langle C_{j}\right\rangle\right\}\rho \right] = 0 \; ,
\end{equation}

\[ Tr\left[\left(C_{j}- \left\langle C_{j}\right\rangle\right)\rho\ln\rho\right]%
- (i/\hbar\sigma)\;Tr\left[\;\left[C_{j}-\left\langle C_{j}\right\rangle,%
\;H - \textrm{E}\right]\;\rho\;\right] + \]

\begin{equation}
 \;\;\;\;\;  +\; \zeta \;Tr\left[\left\{ C_{j} -
\left\langle C_{j} \right\rangle ,H - \textrm{E} \right\}\;\rho\;
\right] + \eta ^{l}\;Tr\left[\left\{ C_{j} - \left\langle C_{j}
\right\rangle ,\;C_{l} - \left\langle C_{l} \right\rangle
\right\}\;\rho\; \right] = 0 \; ,
\end{equation}

\[j = 1,\;2,\ldots n \]

\end{mathletters}

\noindent and we have identified $\sigma(Im\zeta) = (1/\hbar)$,
$Re\zeta \to \zeta $, $\xi  =  - \left[ {{\rm Tr}\left( {\rho \ln
\rho } \right) + 2\left( {{\mathop{\rm Re}\nolimits} \zeta }
\right){\rm E} + 2\eta ^j \left\langle {C_j } \right\rangle }
\right]$. Eqs.(\ref{eq92}) always have solution, as the matrix of
coefficients for the unknowns $\zeta$, $\eta^{j}$ is recognized
to be the positively defined covariance matrix for the
hamiltonian and the operators $C_{j} $. If we presume the
invariance of the hamiltonian as defined by Eq.(\ref{eq83}), in
accordance with the discussion above, it is now straightforward
to verify that $\bar{D}=\zeta\;(H -\textrm{E}) +
\eta^{j}\;\left(C_{j} - \left\langle C_{j} \right\rangle\right)$
is invariant as well in the sense of Eq.(\ref{eq84}a), provided
$\zeta$, $\eta^{j}$ change as

\begin{mathletters}
\label{eq93}

\begin{equation}
 \zeta' = \zeta  \; ,
\end{equation}

\begin{equation}
 \eta'^{j}= \eta^{l}c_{l}^{\;j}+ \zeta\;a^{j} \;.
\end{equation}

\end{mathletters}

\noindent In deriving Eqs.(\ref{eq93}) use has been made of
Eqs.(\ref{eq86}) and the following transformation of $\left\langle
C_{j}\right\rangle$ under the action of $U(\textrm{t})$:

\begin{equation}
\label{eq94} \left\langle {C_{j}} \right\rangle \equiv Tr\left[
{C_{j} \rho}  \right] = Tr\left[ {U\left( {t} \right)C_{j} U^{ +}
\left( {t} \right){\rho} '} \right] = c_{j}^{\;\;l} \left\langle
{C_{j}}  \right\rangle ^{\prime}  + f_{j} \; .
\end{equation}

The complete invariance of Eq.(\ref{eq91}) requires, of course,
that ${\zeta }',\;{\vec {\eta} }'$ defined in Eqs.(\ref{eq93}) be
solutions of the transformed Eqs.(\ref{eq92}), obtained upon
substituting ${\rho} ',{E}'\;and\;\left\langle {\vec {P}}
\right\rangle ^{\prime} $ for $\rho ,E\;and\;\left\langle {\vec
{P}} \right\rangle $, respectively. But substitution of $\rho
\left( {\rm t} \right) = U\left( {\rm t} \right)\rho '\left( {\rm
t} \right)U^ +  \left( {\rm t} \right)$, followed by
rearrangement of \textit{U}, $U^{+}$ over observables and use of
the relations

\begin{mathletters}
\label{eq95}

\begin{equation}
 U\left( {t} \right)HU^{ +} \left( {t} \right) - E = H - {E}' + a^{j}\left(
{C_{j} - \left\langle {C_{j}}  \right\rangle ^{\prime}}\right)\; ,
\end{equation}

\begin{equation}
 U\left( {t} \right)C_{j} U^{ +} \left( {t} \right) - \left\langle {C_{j}}
\right\rangle = c_{j}^{\;\;l} \left( {C_{l} - \left\langle {C_{l}}
\right\rangle}  \right) \; ,
\end{equation}

\[j = 1,\;2,\ldots n \]

\end{mathletters}

\noindent obtained from Eqs.(\ref{eq83}), (\ref{eq86}),
(\ref{eq87}) and (\ref{eq94}), leads to

\begin{mathletters}
\label{eq96}

 \[Tr\left[ {\left( {H - {E}'} \right){\rho} 'ln{\rho} '} \right] + 2\;\zeta
\;Tr\left[ {\left( {H - {E}'} \right)^{2}{\rho} '} \right] +
\left( {\eta ^{l}c_{l}^{\;j} + \zeta \;a^{j}} \right)Tr\left[
{\left\{ {H - {E}',C_{j} - \left\langle {C_{j}} \right\rangle
^{\prime} } \right\}{\rho} '} \right]\]

 \begin{equation}
  + (i/\hbar\sigma)\;a^{j}\;Tr\left[ {\left[ {C_{j} - \left\langle {C_{j}}
\right\rangle^{\prime} ,H - {E}'} \right]{\rho}'}\right] = 0 \; ,
\end{equation}

 \[Tr\left[ {\left( {C_{j} - \left\langle {C_{j}}  \right\rangle ^{\prime} }
\right){\rho} 'ln{\rho} '} \right] - (i/\hbar \sigma)\;Tr\left[
{\left[ {C_{j} - \left\langle {C_{j}}  \right\rangle ^{\prime} ,H
- {E}'} \right]{\rho} '} \right] + \]

\[+ \zeta \;Tr\left[
{\left\{ {C_{j} - \left\langle {C_{j} } \right\rangle ^{\prime}
,H - {E}'} \right\}{\rho} '} \right] + \;\left( {\eta
^{m}c_{m}^{\;\;l} + \zeta a^{l}} \right)Tr\left[ {\left\{ {C_{j}
- \left\langle {C_{j}}  \right\rangle ^{\prime} ,C_{l} -
\left\langle {C_{l}} \right\rangle ^{\prime} } \right\}{\rho} '}
\right] - \]

\begin{equation}
  - (i/\hbar \sigma)\;a^{l}Tr\left[ {\left[
{C_{j} - \left\langle {C_{j}}  \right\rangle ^{\prime} ,C_{l} - \left\langle
{C_{l}}  \right\rangle ^{\prime} } \right]{\rho} '} \right] = 0 \; ,\\
\end{equation}
\[j = 1,\;2,\ldots n \;.\]

\end{mathletters}

\noindent
The first of these equations displays the required
invariance only if the last term vanishes identically, which
demands

\begin{equation}
\label{eq97} \left[ {C_{j} \left( {t} \right),H} \right] = 0\;,
\end{equation}

\noindent
for $j = 1,\;2,\ldots n$, while the second equation is
seen to be invariant provided

\begin{equation}
\label{eq98}
\left[ {C_j ({\rm t}),C_l ({\rm t})} \right] = 0\;,
\end{equation}

\noindent
for $j = 1,\;2,\ldots n$, $l = 1,\;2,\ldots n$. We
conclude that Eq.(91) is invariant if and only if
Eqs.(\ref{eq83}), (\ref{eq97}) and (\ref{eq98}) are
simultaneously verified, in which case the parameters $\zeta $,
$\eta^{j}$ transform according to Eqs.(\ref{eq93}). The
generating variational principle, Eq.(\ref{eq89}), is invariant,
of course, under the same conditions.

Let us substitute now for $U(\textrm{t})$ the special Galilei
boost of velocity $\vec{v}_{0}$,

\begin{equation}
\label{eq99} U\left( {t;\vec {v}_{0}}  \right) = \exp\left[
{\frac{{i}}{{\hbar} }\left( {\vec {P} \cdot t - m \cdot \vec {X}}
\right) \cdot \vec {v}_{0}}  \right]\;,
\end{equation}

\noindent where m is the total mass of the system, $\vec{X}$ is
the position of the center of mass and $\vec{P}$ denotes the
total momentum. Expression (\ref{eq99}) obviously prompts the
identifications $ C_{j}=P_{j}$, $a^{j}=1$, $b=(m\vec{v}_{0})/2$,
$c_{j}^{\;\;l}=\delta _{jl}$, $f_{j} = m\vec {v}_{0}$, which
introduced in Eqs.(\ref{eq97}) and (\ref{eq98}) lead to the
recognizable commutation relations

\begin{equation}
\label{eq100} \left[ {H,P_{j}} \right] = \left[ {P_{j} ,P_{l}}
\right] = 0\;.
\end{equation}

\noindent
Subsequent substitution in Eq.(91) gives the
corresponding equation of motion in the form

\begin{equation}
\label{eq101} \dot {\rho}  = - \sigma \left[ {\rho \ln\rho +
\left\{ {\zeta \left( {H - E} \right) + \eta ^{j}\left( {P_{j} -
\left\langle {P_{j}}  \right\rangle} \right),\rho}  \right\} -
\rho \;Tr\left( {\rho \ln\rho}  \right)} \right] +
\frac{{i}}{{\hbar} }\left[ {\rho ,H} \right]\;.
\end{equation}

\noindent
We recover thus in an unexpected manner the celebrated
result that the Galilei-invariance of the appropriate
non-relativistic equation of motion is equivalent to the
corresponding invariance of the hamiltonian, the conservation of
total momentum and the commutation of the hamiltonian and the
total momentum operators.

Eq.(\ref{eq101}) reduces to the original Eq.(\ref{eq24}) in the
center-of-mass referential, where only states corresponding to an
eigenstate of zero total momentum for the center-of-mass
coordinates need be considered and the dissipative momentum terms
vanish. It also retains the fundamental features previously
outlined for Eq.(\ref{eq24}). In particular, it can be checked
that pure states evolve according to the usual hamiltonian
dynamics, the entropy of mixed states increases and the nature of
the asymptotic equilibrium states is preserved, up to a slight
change of form which accounts for the conservation of momentum.
It is also evident that Eq.(\ref{eq101}) is invariant under
time-independent symmetry transformations which leave the
hamiltonian and the dissipator $\bar {D}$ invariant, provided the
time-scale parameter $\sigma $ has the same property. In
particular, if the hamiltonian commutes with the total angular
momentum, Eq.(\ref{eq101}) is invariant under finite rotations.
However, as for the linear momentum, rotational invariance alone
does not imply, in general, a conservation law for the angular
momentum. The latter can be brought into view by requiring that
the equation of motion for the density matrix be covariant with
respect to all reference frames where the conservation of energy
is a valid physical law. In particular we should consider
translations to observers in uniform rotational motion around an
axis at rest in some inertial frame. The rather cumbersome
details of adding this supplementary constraint will be left
aside, since nothing new will be gained for the formalism.

A more interesting lack of symmetry for Eq.(\ref{eq101}), or
better, the simpler Eq.(\ref{eq24}), lies concealed in the
apparent absence of separability. Indeed, let the system
described by the hamiltonian \textit{H} be composed of two
noninteracting subsystems, such that $H = H_{1}+H_{2}$,
$[H_{1},\;H_{2}] = 0$, and consider the situation of a separable
initial state $\rho(0)=\rho_{1}(0)\rho_{2}(0)$, of energy
$\textrm{E} = \textrm{E}_{1}+\textrm{E}_{2}$. Direct inspection of
Eq.(\ref{eq24}) shows that the energies of the two subsystems
cannot be separately conserved and a completely separable
solution is thus prohibited. But we also observe that lifting the
constraint of separate conservation of energy allows a
pseudo-separable solution
$\rho(\textrm{t})=\rho_{1}(\textrm{t})\;\rho_{2}(\textrm{t})$
given by the coupled system

\begin{mathletters}
\label{eq102}

\begin{equation}
\dot{\rho}_{1}=-\sigma\;\left[\rho _{1}\ln\rho_{1}+\zeta\;
\left\{H_{1} -
\frac{Tr\left(H_{1}\rho_{1}\right)}{Tr\left(\rho_{1}\right)}\;,\;\rho_{1}\right\}
- \rho_{1}\;\frac{Tr\left( {\rho _{1} \ln\rho _{1}}
\right)}{Tr\left( {\rho _{1}} \right)} \right] +
\frac{i}{\hbar}\left[\rho _{1},\;H_{1}\right]\;,
\end{equation}

\begin{equation}
 \dot{\rho}_{2} =
- \sigma\;\left[\rho_{2}\ln\rho_{2}+ \zeta\;
\left\{H_{2}-\frac{Tr\left(H_{2} \rho
_{2}\right)}{Tr\left(\rho_{2}\right)}\;,\;\rho_{2}\right\} -
\rho_{2} \;\frac{Tr\left(\rho_{2}\ln\rho_{2}\right)}{Tr\left(\rho
_{2}\right)}\right] +
\frac{i}{\hbar}\left[\rho_{2},\;H_{2}\right]\;.
\end{equation}

\end{mathletters}

\noindent In this case probability is independently conserved for
each subsystem, since $Tr(\dot{\rho}_{i})=0$, while energy is only
conserved globally,

 \[\frac{Tr\left(H_{1}\rho
_{1}\right)}{Tr\left(\rho_{1}\right)} + \frac{Tr\left(
H_{2}\rho_{2}\right)}{Tr\left(\rho_{2}\right)} = \textrm{E}\;.\]

\noindent The coupling between the (noninteracting) subsystems
appears to be as instantaneous and nonlocal as usual quantum
entanglement, but unlike the latter, it involves an unorthodox
exchange of energy. The significance of this unusual outcome
follows from the observation that, according to
Eqs.(\ref{eq102}), the equilibrium of the compound system is
attained for values of $\sigma$ and $\zeta$ common to both
subsystems, hence for a common generalized temperature. Imagine
now that the initial states for the two subsystems are chosen as
individual equilibrium states with different corresponding
temperatures. It follows that the dynamics given by
Eq.(\ref{eq24}) will drive the total system towards a new state
of equilibrium, with a temperature common to both components. We
cannot but concede the obvious similarity of this unconventional
effect with the classical process of equilibration by thermal
contact. Its origin lies in the very assumption of maximal
entropy increase on which Eq.(\ref{eq24}) has been derived.
Indeed, even when the entropy of each subsystem is already
maximal under individual isolation, if states of larger total
entropy are available, probabilities and energy (heat) will be
necessarily redistributed so as to enforce a further increase of
the overall entropy. Whether this entropic entanglement, or ideal
thermal contact, is or not an element of reality appears
equivalent to accepting or rejecting the conjecture that an
isolated, perfectly ideal gas can undergo relaxation towards
equilibrium.

We can provide formal support towards the positive by pointing out that the
effect of entropic entanglement does not necessarily interfere with the
concept of separable evolution for mutually isolated systems. First let us note that explicitly specifying an adiabatic separation (in the thermodynamic sense) of the noninteracting systems, and hence allowing for separate conservation of energy, removes most of the entropic entanglement. In this case the resulting equation of motion will display distinct $\zeta$-s for each of the systems, but a common time-scale parameter, i.e. 

\begin{equation}
\label{eq103} 
\dot {\rho}  = - \sigma \left[ {\rho \ln\rho +
 \zeta_{1} \left\{H_{1} - E_{1},\rho \right\}+
 \zeta_{2} \left\{ H_{2} - E_{2},\rho \right\} -
\rho \;Tr\left( {\rho \ln\rho}  \right)} \right] +
\frac{{i}}{{\hbar} }\left[ {\rho ,H_{1}+H_{2}} \right]\;,
\end{equation}

\noindent 
with

\[E_{i}=\frac{Tr(H_{i}\rho)}{Tr(\rho)}=const.\;,\]

\noindent for $i=1, 2$. As before, it proves possible to extract a pseudo-separable solution $\rho(t)=\rho_{1}(t)\rho_{2}(t)$, but Eqs.(\ref{eq102}) are replaced by 

\begin{mathletters}
\label{eq104}

\begin{equation}
\dot{\rho}_{1}=-\sigma\;\left[\rho _{1}\ln\rho_{1}+\zeta_{1}
\left\{H_{1} - E_{1}\;,\;\rho_{1}\right\} - \rho_{1}\;\frac{Tr\left( {\rho _{1} \ln\rho _{1}}
\right)}{Tr\left( {\rho _{1}} \right)} \right] +
\frac{i}{\hbar}\left[\rho _{1},\;H_{1}\right]\;,
\end{equation}

\begin{equation}
 \dot{\rho}_{2} =
- \sigma\;\left[\rho_{2}\ln\rho_{2}+ \zeta_{2}
\left\{H_{2}-E_{2}\;,\;\rho_{2}\right\} -
\rho_{2} \;\frac{Tr\left(\rho_{2}\ln\rho_{2}\right)}{Tr\left(\rho
_{2}\right)}\right] +
\frac{i}{\hbar}\left[\rho_{2},\;H_{2}\right]\;.
\end{equation}

\end{mathletters}

\noindent 
where this time the $\zeta_{i}$ parameters, $i=1, 2$,  will be found to depend only on the corresponding $\rho_{i}$ and $H_{i}$, in exactly the manner obtained for a single isolated system. Yet the two evolutions remain tethered by the time-scale parameter $\sigma$, thus retaining a weaker form of entropic entanglement. The simple presence of other noninteracting, adiabatically separated systems appears to alter the time-scale of dissipative relaxation for any given system. If $\sigma$ is assumed variable in time, e.g. through a dependence on $\rho$, this influence will be time-dependent unless all other systems have reached equilibrium. But since $\sigma$ does not affect the nature of the asymptotic equilibrium state, the equilibrium of any one system will not be disturbed by other systems and will display an individual temperature determined solely by the corresponding energy content. 

A careful examination will trace the above type of nonseparability to the fact that the corresponding variational principle selects the direction of maximum entropy increase by referring to the time derivative of the total (entangled) state operator, and not to disentangled, individual state operators separately. However, this pitfall can be avoided if it is recognized that true mutual isolation precludes entanglement on invariance grounds. Indeed, regardless of the nature of the underlying dynamics, the evolution of two mutually isolated systems should remain invariant under \textit{every} transformation pertaining to the individual symmetry groups. In particular, it should be invariant under individual time translations. Since entangled states certainly do not possess this invariance, they do not describe truly isolated systems. In other words, the restricted subspace of the state space that can be spanned
by the dynamics of mutually isolated systems should contain only non-entangled states and the evolution of each of the factor states should be driven independently. In our nonlinear setting, where this subspace is selected by means of the generating variational principle, this restriction has to be correctly built in the variational functional itself. Hence one has to account both for individual conservation laws, excluding thus any energy exchange, as well as for vanishing entanglement. The latter imposes a separable state operator $\gamma(t)=\gamma_{1}(t)\gamma_{2}(t)$ and also requires that the entropy production be maximized separately with respect to variations of $\dot{\gamma}_{1}$ and $\dot{\gamma}_{2}$,  i.e. the $\sigma$ term in the variational principle should be replaced according to

\[\frac{2}{\sigma}(\dot{\gamma}|\dot{\gamma})\;\;\rightarrow\;\;%
\frac{2}{\sigma_{1}}(\dot{\gamma}_{1}\gamma_{2}|\dot{\gamma}_{1}\gamma_{2})+%
\frac{2}{\sigma_{2}}(\gamma_{1}\dot{\gamma}_{2}|\gamma_{1}\dot{\gamma}_{2}) \]

\noindent with each  $\sigma_{i}$ a functional only of $\gamma_{i}$ and $H_{i}$. But then the variational principle takes the form

\begin{mathletters}
\label{eq105}

\begin{equation}
\delta\{\left(\gamma_{2}|\gamma_{2}\right)F_{1}+\left(\gamma_{1}|\gamma_{1}\right)F_{2}\}\;=\;0\;\;,
\end{equation}

\[ F_{i}=\;(\dot{\gamma}_{i}|\ln(\gamma_{i}\gamma_{i}^{+})|\gamma_{i})+%
(\gamma_{i}|\ln(\gamma_{i}\gamma_{i}^{+})|\dot{\gamma}_{i})+%
2\;\zeta_{i}(\dot{\gamma}_{i}|\textbf{H}_{i}|\gamma_{i})+%
2\;\zeta_{i}^{*}(\gamma_{i}|\textbf{H}_{i}|\dot{\gamma}_{i}) +
 \]

\begin{equation}
+\left[\bar{\xi}_{i}(\dot{\gamma}_{i}|\gamma_{i})+\bar{\xi}_{i}^{*}(\gamma_{i}|\dot{\gamma}_{i})\right]+%
\frac{2}{\sigma_{i}}(\dot{\gamma}_{i}|\;\dot{\gamma}_{i})\;\;, 
\end{equation}

\[i=1,\;2\;\;,\]

\end{mathletters}

\noindent where

\[\bar{\xi}_{1}=\xi_{1}+\left( \xi_{2}+\zeta_{2}\rm{E}_{2}-\frac{\rm{S}_{2}}{\rm{k}_{\rm{B}}(\gamma_{2}|\gamma_{2})}\right )\]
\[\bar{\xi}_{2}=\xi_{2}+\left( \xi_{1}+\zeta_{1}\rm{E}_{1}-\frac{\rm{S}_{1}}{\rm{k}_{\rm{B}}(\gamma_{1}|\gamma_{1})}\right )\]

\noindent Independent variation on $\dot{\gamma}_{1}$ and $\dot{\gamma}_{2}$, followed by extraction of the Lagrange parameters from the corresponding conservation conditions leads now to the desired separate equations of motion for $\rho_{i}=\gamma_{i}\gamma_{i}^{+}$,

\begin{equation}
\label{eq106}
\dot{\rho}_{i}=-\sigma_{i}\left[\;\rho_{i}\ln\rho_{i} +%
\zeta_{i}\{H_{i} - \textrm{E}_{i},\;\rho_{i}\}-%
\rho_{i}\frac{Tr(\rho_{i}\ln\rho_{i})}{Tr(\rho_{i})}\right]+\frac{i}{\hbar}\;[\rho_{i},\;H_{i}]\;, 
\end{equation}

\[i=1,\;2\;\;,\]

\noindent  with 

\[\sigma_{i}=\sigma_{i}(\rho_{i},H_{i}) \ge 0\;,\]
\[\zeta_{i}=-\frac{1}{2}\;\frac{Tr[(H_{i} -\rm{E}_{i})\rho_{i}\ln\rho_{i}]}%
{Tr[(H_{i} -\rm{E}_{i})^2\rho_{i}]}\;,\]
\[\textrm{E}_{i}=\frac{Tr(H_{i}\rho_{i})}{Tr(\rho_{i})}=const.\;.\]

\noindent Obviously, the invariance of the nonlinear dynamics under the symmetry group of each component subsystem is so restored, provided the $\sigma_{i}$-s are invariant also.

\section{GENERALIZATION TO ARBITRARY ENTROPY AND ENERGY FUNCTIONAL FORMS}
\label{sec6}

The framework developed in the previous Secs. can be easily expanded to accommodate non-standard entropy functionals and/or energy forms with a nonlinear dependence on the density matrix $\rho$. This generalized formalism can then provide nonlinear extensions for, e.g., the Lie-Poisson dynamics or a standard hamiltonian evolution supplemented by a nonextensive Tsallis entropy \cite{9}, appropriate for systems with fractal properties. We sketch here only the derivation of the generalized equation of motion, since a detailed analysis exceeds the purpose of the present work.

To this end, let us start with a Lie-Poisson equation of motion in the form \cite{6}

\begin{equation}
\label{eq107}
\dot{\rho} = -\frac{i}{\hbar}\left[\rho, \;\hat{H}(\rho)\right]
\end{equation}

\noindent where $\hat{H}(\rho)$ is in general a hermitian, nonlinear functional of $\rho$. The energy conservation law is now replaced by

\[Tr\left(\hat{H}(\rho)\dot{\rho}\right) = 0\] 

\noindent or in terms of the state operator $\gamma$,

\begin{equation}
\label{eq108}
(\dot{\gamma}|\hat{\bf{H}}(\rho)|\gamma)+(\gamma|\hat{\bf{H}}(\rho)|\dot{\gamma}) = 0
\end{equation}

\noindent The law of probability conservation, on the other hand, remains unchanged since $Tr(\dot{\rho})=0$ or

\begin{equation}
\label{eq109}
(\dot{\gamma}|\gamma)+(\gamma|\dot{\gamma}) = 0
\end{equation}

 Let us search now for a nonlinear evolution that observes the above conservation constraints, Eqs.(\ref{eq108}) and (\ref{eq109}), and is also subject to a second principle based on some unspecified, positive definite entropy functional
 $\frac{\textrm{S}}{\textrm{k}_{\rm{B}}} = Tr(\hat{S}(\rho))$, such that $\dot{\textrm{S}} = Tr\left((\delta \hat{S}/\delta \rho)\dot{\rho}\right) \ge 0$ or 

\begin{equation}
\label{eq110}
(\dot{\gamma}|\frac{\delta \hat{\bf{S}}}{\delta \rho}|\gamma)+(\gamma|\frac{\delta \hat{\bf{S}}}{\delta \rho}|\dot{\gamma}) \ge 0
\end{equation}

\noindent Here the operator $\hat{S}(\rho)$ is assumed hermitian and $(\delta \hat{S}/\delta \rho)$ denotes its hermitian functional derivative with respect to $\rho$. The corresponding variational principle is now written

\begin{equation}
\label{eq111} 
\delta \left\{  -(\dot{ \gamma} \;
|\frac{\delta \hat{\bf{S}}}{\delta \rho}|\gamma)-(\gamma|\frac{\delta \hat{\bf{S}}}{\delta \rho}|\dot{\gamma})+
2\;\zeta\;(\dot{\gamma}|\hat{\bf{H}}(\rho)|\gamma)+2\;\zeta^{*}(\gamma|\hat{\bf{H}}(\rho)|\dot{\gamma})+%
\xi\;[\;(\dot{\gamma}|\gamma)+(\gamma|\dot{\gamma})\;]+\;\frac{2}{\sigma}\;(\dot{\gamma}|\dot{\gamma})\; \right\} = 0\;.
\end{equation}

\noindent and can be verified to generate the following equation of motion:

\begin{equation}
\label{eq112}
\dot{\rho}=-\sigma\left[\;-\frac{\delta \hat{S}}{\delta \rho}\rho +%
\zeta\;\left\{ \hat{H}(\rho) -\left\langle\hat{H}(\rho)\right\rangle,\;\rho \right\}+%
\left\langle \frac{\delta \hat{S}}{\delta \rho} \right\rangle \rho\;\right]+\frac{i}{\hbar}\;\left[\rho,\;\hat{H}(\rho)\right]\;,
\end{equation}

\noindent where

\[ \left\langle A \right\rangle = \frac{Tr(A\rho)}{Tr(\rho)} \;,\]

\noindent and 

\[ \zeta =  \frac{1}{2} \frac{ \left\langle \left( \hat{H}(\rho)-\left\langle \hat{H}(\rho) \right\rangle\right)%
(\delta \hat{S}/ \delta \rho)\right\rangle}%
{ \left\langle \left( \hat{H}(\rho)-\left\langle \hat{H}(\rho) \right\rangle\right)^{2}\right\rangle } \;\;,\]

\[ \sigma = \sigma \left( \rho, \hat{H}(\rho)-\left\langle \hat{H}(\rho) \right\rangle \right) \ge 0 \;\;.\]

\noindent We note that if $(\delta \hat{S}/\delta \rho) \rho = 0 $ for pure states, $\rho = \rho^{2}$, then $\langle \delta \hat{S} / \delta \rho \rangle = 0$, $\zeta = 0$, and the pure state dynamics reduces to that prescribed by Eq.(\ref{eq107}). 

When the energy functional reduces to the hamiltonian, $\hat{H}(\rho) = H $, and the entropy is given the standard von Neumann expression, such that $\hat{S}(\rho) = -\rho\ln\rho$, $(\delta \hat{S} / \delta \rho) \rho = -\rho\ln\rho-\rho$,  we recover the basic Eq.(\ref{eq24}). A $\rho$-dependent $\hat{H}(\rho)$, complemented by the standard entropy, leads to a nonlinear extension of the Lie-Poisson dynamics,

\begin{equation}
\label{eq113}
\dot{\rho}=-\sigma\left[\;\rho\ln\rho +%
\zeta\;\left\{\hat{H}(\rho) -\left\langle\hat{H}(\rho)\right\rangle,\;\rho\right\}-%
\frac{Tr(\rho\ln\rho)}{Tr(\rho)} \rho\;\right]+\frac{i}{\hbar}\;\left[\rho,\;\hat{H}(\rho)\right]\;,
\end{equation}

\noindent with 

\[ \zeta = - \frac{1}{2} \frac{Tr\left[\left(\hat{H}-\left\langle \hat{H}(\rho) \right\rangle\right)\rho\ln\rho\right]}%
{Tr\left[ \left( \hat{H}(\rho)-\left\langle \hat{H}(\rho) \right\rangle\right)^{2}\rho\right]} \;\;.\]

\noindent If $\hat{H}(\rho)$ is reduced to the standard hamiltonian $H$, but the entropy is given a Tsallis form, with

\[\hat{S}(\rho) = - \frac{\rho-\rho^{q}}{q-1}\;,\]
\[\frac{\delta \hat{S}}{\delta \rho} \rho =\frac{\rho-q\rho^{q}}{q-1}\;,\]

\noindent for given real q, the result will be a nonlinear extension of the von Neumann dynamics under Tsallis q-thermostatistics, which reads, after a few elementary manipulations,

\begin{equation}
\label{eq114}
\dot{\rho}=-\sigma\left[\;\frac{q}{q-1}\rho^{q} +%
\zeta\;\{H -E,\;\rho\}-\frac{q}{q-1}\frac{Tr\left(\rho^{q}\right)}{Tr(\rho)} \rho\;\right]+\frac{i}{\hbar}\;[\rho,\;H]\;,
\end{equation}

\noindent where

\[\zeta = -\frac{1}{2}\frac{q}{q-1}\frac{Tr\left[(H-\rm{E})\rho^{q}\right]}{Tr\left[(H-\rm{E})^2\rho\right]} \]

\noindent Situations where the standard averages have to be replaced by q-averages can be approached in the same fashion, by appropriately redefining the conserved functionals.

\section{CONCLUSION}
\label{sec7}

We have constructed and analyzed a non-relativistic nonlinear
extension of the quantum law of evolution, which accounts for the
second principle of thermodynamics \textit{and} is not at odds with the
factual linearity of pure state propagation. The theoretical
existence of such an extension confirms that the linear and
unitary evolution of pure states is not in itself sufficient
proof for the general linearity of quantum mechanics \cite{5,6}.
One must conclude that the linear propagation of mixed states also
has to be corroborated experimentally, to comparable precision,
before a definitive conclusion can be drawn. It is hoped that the
formal study developed here provides a meaningful benchmark in
this sense.

Our main result is  Eq.(\ref{eq24}), which defines the modified
time evolution of the density matrix. The equation of motion was
extracted from a variational principle on the space of state
operators, rather than the space of density matrices, as a
trajectory of maximal entropy production under the constraint of
energy and probability conservation, augmented eventually by the
requirement of Galilei invariance (see Eq.(\ref{eq101})). Should
we drop the requirement of entropy increase, the parameters
$Re\zeta,\;\xi$ vanish and the equation of motion reduces
automatically to the common hamiltonian form. The outlined
procedure may not be unique, but is encouraging in its
consistency. In addition, it applies as well to alternate theories which use nonstandard energy or entropy forms.  It is notable that the variational principle has
sense only in terms of state operators, whereas the equation of
motion can be stated simply in terms of the conventional density
matrix.

A peculiar and unexpected idea brought forth in our ansatz is
that a maximal increase of entropy does not necessarily result in
maximal decoherence, to the effect that a pure state of a
perfectly isolated system is not allowed to evolve into a mixed
state. On the contrary, the proposed quantum equivalent of the
second principle of thermodynamics is seen to introduce only a
limited degree of decoherence, in the sense that the cardinality
of the set of nonzero eigenvalues of the density matrix is
preserved. As already mentioned, for the particular case of a
pure initial state this leads to the usual unitary evolution. The
same property also supports, aside from canonical equilibrium
states, a rich class of ``negative-temperature'' equilibrium
states, which bring to mind the notion of thermal coherence.
Furthermore, the ideal thermal contact phenomenon discussed in
Sec.5 abides by the same rule and, according to Eqs.(\ref{eq102}),
a system in an initially pure state will remain in a pure state
even if it is in contact with, but not necessarily interacting
with, other systems. However, in that case the pure state
undergoes relaxation according to a dynamics of Gisin type
\cite{14}, as seen by taking, e.g., $\rho_{1}=\rho_{1}^{2}$,
$\rho_{1}\ln\rho_{1}=0$, in Eq.(\ref{eq102}a),

\begin{equation}
\label{eq115} \dot{\rho_{1}}=-\sigma\zeta\left\{\;H_{1} -
\frac{Tr\left(H_{1}\rho_{1}\right)}{Tr\left(\rho_{1}\right)}\;,\;\rho_{1}\right\}+
\frac{i}{\hbar}\left[\rho _{1},\;H_{1}\right]\;.
\end{equation}

\noindent Depending on the sign of $\zeta$, the asymptotic
stationary state is an energy eigenstate for the lowest (if
$\zeta > 0$) or for the highest (if $\zeta < 0$) energy level
contributing to the initial state $\rho_{1}(0)$. As detailed in
Sec.5, the state of thermal contact is not to be mistaken for a
state of mutual isolation, despite the absence of explicit
interactions.

We find it promising that bending quantum dynamics to account for
classical phenomenological irreversibility suggests a rather
unified picture of both reversibility and irreversibility, as
well as coherence and decoherence, while preserving such
fundamental features as symmetry invariance. However, the
self-consistency of the theory is limited at this point by the
need for an explicit expression for the entropy production, which
means that the equation of motion remains determined up to the
scale setting functional \textit{$\sigma $}. We leave the
resolution of this problem for future consideration, although a
definite expression for \textit{$\sigma $} certainly conditions
the consistency of our results. For instance, Eq.(\ref{eq59}) for
the near-equilibrium damping constants of the density matrix
elements between energy eigenstates shows an acceptable
dependence on the energy gap between the states, but the wrong
temperature dependence ($\gamma _{\mu \nu } \to 0\;as\;\beta \to
0\;$, $\gamma _{\mu \nu}  \to \infty \;as\;\beta \to \infty $) if
\textit{$\sigma $} is assumed temperature independent. In the
least, this observation serves to hint that $\sigma$ should
behave like $\beta^{-(2 + \delta)},\;\delta > 0$, in the vicinity
of canonical equilibrium, which in turn can be used, of course,
as a theoretical benchmark.

\end{document}